\documentclass[11pt, a4paper]{article}

\usepackage{amsmath,amsfonts,bm}
\usepackage{mathrsfs}

\usepackage{amsthm}
\usepackage[left=2cm, right=2cm, bottom=2cm, top=2cm]{geometry}
\usepackage{xcolor}
\usepackage{svg}
\usepackage{booktabs}
\usepackage{graphicx}
\usepackage{caption}
\usepackage{subcaption}

\usepackage{url}

\definecolor{amethyst}{rgb}{0.6, 0.4, 0.8}
\usepackage{authblk}
\usepackage[colorlinks, citecolor=blue, linkcolor=teal, urlcolor=amethyst]{hyperref}
\usepackage[square, numbers, comma, sort&compress]{natbib}
\usepackage[capitalise, nameinlink]{cleveref}

\title{DiffPaSS -- High-performance differentiable pairing of protein sequences using soft scores}

\author{Umberto Lupo\textsuperscript{1,2,$\dagger$}, Damiano Sgarbossa\textsuperscript{1,2}, Martina Milighetti\textsuperscript{3,4}, Anne-Florence Bitbol\textsuperscript{1,2,$\dagger$}}
\affil{\textbf{1} Institute of Bioengineering, School of Life Sciences, École Polytechnique Fédérale de Lausanne (EPFL), CH-1015 Lausanne, Switzerland\\
\textbf{2} SIB Swiss Institute of Bioinformatics, CH-1015 Lausanne, Switzerland\\
\textbf{3} Division of Infection and Immunity, University College London, London, United Kingdom \\
\textbf{4} Cancer Institute, University College London, London, United Kingdom\\
$^\dagger$ Emails: \href{mailto:umberto.lupo@epfl.ch}{\texttt{umberto.lupo@epfl.ch}}, \href{mailto:anne-florence.bitbol@epfl.ch}{\texttt{anne-florence.bitbol@epfl.ch}}}
\date{}

\begin{document}

\maketitle

\begin{abstract}
Identifying interacting partners from two sets of protein sequences has important applications in computational biology. Interacting partners share similarities across species due to their common evolutionary history, and feature correlations in amino acid usage due to the need to maintain complementary interaction interfaces. Thus, the problem of finding interacting pairs can be formulated as searching for a pairing of sequences that maximizes a sequence similarity or a coevolution score. Several methods have been developed to address this problem, applying different approximate optimization methods to different scores. We introduce DiffPaSS, a differentiable framework for flexible, fast, and hyperparameter-free optimization for pairing interacting biological sequences, which can be applied to a wide variety of scores. We apply it to a benchmark prokaryotic dataset, using mutual information and neighbor graph alignment scores. DiffPaSS outperforms existing algorithms for optimizing the same scores. We demonstrate the usefulness of our paired alignments for the prediction of protein complex structure. DiffPaSS does not require sequences to be aligned, and we also apply it to non-aligned sequences from T cell receptors.
\end{abstract}

\section*{Introduction}
\label{introduction}

Identifying which proteins interact together, using their sequence data alone, is an important and combinatorially difficult task.
Mapping the network of protein-protein interactions, and predicting the three-dimensional structures of individual protein complexes, often requires determining which sequences are functional interaction partners among the paralogous proteins of two families. In the case of specific one-to-one interactons, this problem can be formulated as looking for a permutation of the sequences one family with respect to those of the other within each species.
Sequence similarity-based scores~\cite{goh2002co,Ramani03,Gertz03,Izarzugaza06,tillier2006codep,Izarzugaza08,Tillier09,Bradde10,Hajirasouliha12,ElKebir13} and coevolution-based ones~\cite{Burger08,Bitbol16,Gueudre16,Bitbol18}, as well as a method combining both ingredients~\cite{Gandarilla23} have been proposed to tackle this problem. 

Sequence similarity is employed because when two proteins interact in one species, and possess close homologs in another species, then these homologs are likely to also interact. More generally, interacting protein families share a similar evolutionary history~\cite{Pazos01,Ochoa10,Ochoa15}, leading to sequence similarity-based pairing methods~\cite{goh2002co,Ramani03,Gertz03,Izarzugaza06,tillier2006codep,Izarzugaza08,Tillier09,Bradde10,Hajirasouliha12,ElKebir13}. The use of neighbor graph alignment~\cite{Bradde10} or of orthology determined by closest reciprocal hits for pairing interaction partners~\cite{cong2019protein,Green21,evans2021protein,Humphreys2021} also relies on this idea. 

The idea underlying coevolution-based methods for pairing interacting sequences is that amino acids that are in contact at the interface between two interaction partners need to maintain physico-chemical complementarity through evolution, which gives rise to correlations in amino-acid usage between interacting proteins~\cite{Burger08,Weigt09,Bitbol16,Gueudre16,Bitbol18}. Mutual information~\cite{Dunn08} and pairwise maximum entropy models~\cite{Weigt09,Schug09,Morcos11,Marks11,Sulkowska12,Mann14} can reveal such coevolution both within a protein sequence and between the sequences of interacting partners. Additional correlations come from the shared evolutionary history of interacting partners~\cite{Marmier19,Gerardos22}. Thus, permutations maximizing coevolution scores are expected to encode correct interactions \citep{Bitbol16,Bitbol18,Gandarilla23}. Coevolution-based approaches require large and diverse multiple sequence alignments (MSAs) to perform well, which limits their applicability. 
More recently, scores coming from protein language models have also been proposed~\cite{Chen23,Lupo2024}. In particular, we proposed a method that outperforms traditional coevolution-based methods for shallow MSAs, comprising few sequences. However, this method is computationally intensive, and memory requirements limit its applicability to large MSAs.

We present DiffPaSS (``\textbf{Diff}erentiable \textbf{Pa}iring using \textbf{S}oft \textbf{S}cores''), a family of flexible, fast and hyperparameter-free algorithms for pairing interacting sequences among the paralogs of two protein families. DiffPaSS optimizes smooth extensions of coevolution or similarity scores to ``soft'' permutations of the input sequences, using gradient methods. It can be used to optimize any score, including coevolution scores and sequence similarity scores.
Strong optima are reached thanks to a novel bootstrap technique, motivated by heuristic insights into this smooth optimization process.
When using inter-chain mutual information (MI) between two MSAs as the score to be maximised, DiffPaSS outperforms existing coevolution- and sequence similarity-based pairing methods on difficult benchmarks composed of small MSAs from ubiquitous interacting prokaryotic systems.
Compared to the protein language model based method DiffPALM~\cite{Lupo2024}, DiffPaSS  more rapidly produces paired alignments that can be used as input to AlphaFold-Multimer \citep{evans2021protein}, in order to predict the three-dimensional structure of protein complexes.
We show promising results in this direction, for some eukaryotic complexes where the default AlphaFold-Multimer settings do not provide good performance. 
DiffPaSS is a general method that is not restricted to coevolution scores. In particular, it can be used to pair sequences by aligning their similarity graphs. We demonstrate that DiffPaSS outperforms a Monte Carlo simulated annealing method for graph alignment-based pairing of interacting partners on our benchmark prokaryotic data.
Importantly, DiffPaSS graph alignment can be used even when reliable MSAs are not available. We show that it outperforms Monte Carlo graph alignment on the problem of pairing non-aligned sequences of T-cell receptor (TCR) CDR3$\alpha$ and CDR3$\beta$ loops. 
A \texttt{PyTorch} implementation and installable Python package are available at \url{https://github.com/Bitbol-Lab/DiffPaSS}.

\section*{Methods}
\label{methods}

\subsection*{Problem and general approach}

\paragraph{Pairing interacting protein sequences.} Consider two protein families A and B that interact together, and the associated MSAs $\mathcal{M}_\mathrm{A}$ and $\mathcal{M}_\mathrm{B}$. Each of them is partitioned into $K$ species, and we denote by $N_k$ the number of sequences in species $k$, with $k = 1, \ldots, K$. In practice, the number of members of family A and of family B in a species is often different. In this case, $N_k$ is the largest of these two numbers, and we add sequences made entirely of gap symbols to the MSA with fewer sequences in that species, as in \citet{Lupo2024}.

Our goal is to pair the proteins of family A with their interacting partners from family B within each species. We assume for simplicity that interactions are one-to-one within each species. This is the first and main problem addressed by our method, and is commonly known as the paralog matching problem, as the within-species protein variants are often paralogs.

Our method also extends to related matching problems involving non-aligned sequences. This will be discussed below, with an application to variable regions of T-cell receptors, where alignment is challenging. Then, $\mathcal{M}_\mathrm{A}$ and $\mathcal{M}_\mathrm{B}$ are ordered collections of non-aligned amino-acid sequences instead of MSAs.

\paragraph{Formalization.} Let $\mathscr{S}$ be a score function of two MSAs (or ordered collections of non-aligned sequences) $\mathcal{M}_\mathrm{A}$ and $\mathcal{M}_\mathrm{B}$, which is sensitive to the relative ordering of the rows of $\mathcal{M}_\mathrm{A}$ with respect to those of $\mathcal{M}_\mathrm{B}$. Our matching problems can be formalized as searching for a permutation $\pi$ of the rows of $\mathcal{M}_\mathrm{A}$ which maximises $\mathscr{S}(\pi(\mathcal{M}_\mathrm{A}), \mathcal{M}_\mathrm{B})$, denoted by $\mathscr{S}(\pi)$ for brevity. The permutation $\pi$ should operate within each species and not across them, since interactions occur within a species. 
There are $\prod_{k=1}^{K} N_k !$ permutations satisfying this constraint, which usually renders unfeasible the brute-force approach of scoring all of them and picking the one with the largest score.

\paragraph{General approach.} Optimizing a score $\mathscr{S}$ across permutations is a discrete problem, but we propose an approximate differentiable formulation.  
Briefly, we first construct a differentiable extension $\hat{\mathscr{S}}$ of $\mathscr{S}$ from permutation matrices to a larger space of matrices.
This larger space comprises square matrices with non-negative entries and whose all rows and columns (approximately) sum to $1$.
In what follows, we refer to such matrices as ``soft permutations'', and to true permutation matrices as ``hard permutations''.
Using real square matrices, referred to as ``parameterization matrices'' $X$, a method \cite{Mena2018} based on the Sinkhorn operator~\citep{sinkhorn1964relationship}
allows to smoothly navigate the space of soft permutations while keeping track of the ``nearest'' hard permutations (for more details, see the Supplementary material).
When this navigation is guided by gradient ascent applied to the extended score function $\hat{\mathscr{S}}$, these hard permutations provide candidate solutions to the original problem.

In general, however, this differentiable optimization problem for $\hat{\mathscr{S}}$ may have several local optima.
Besides, depending on the precise way in which the discrete score $\mathscr{S}$ is extended to a differentiable score $\hat{\mathscr{S}}$ for soft permutations, optimal soft permutations for $\hat{\mathscr{S}}$ may be too distant from any hard permutation to approximate a well-scoring hard permutation for the original problem.
Indeed, we found empirically, for a variety of scores, extensions, and random initializations, that na\"{i}vely applying the procedure above often yielded hard permutations with sub-optimal scores, even after several gradient steps.
Nevertheless, we empirically found the following to be true for a wide class of scores and extensions thereof:
\begin{enumerate}
    \item\label{item:first_grad_step} When gradient ascent is initialized so that the initial soft permutation for each species is ``as soft as possible'' -- i.e., it is a square matrix with all entries equal to the same positive value, and normalized rows and columns -- the first gradient step leads to a nearby \emph{hard} permutation with significantly higher score than given by random expectation. This point is illustrated in \cref{fig:hk-rr_precision_first_gradient_step}.
    \item\label{item:importance_of_ground_truth} Pairing performance increases if some sequences are correctly paired and excluded from gradient optimization, while being included in the calculation of $\mathscr{S}$ and $\hat{\mathscr{S}}$.  As the size of this fixed context is increased, pairing performance for the remaining sequences increases. This point is consistent with previous approaches using other methods \citep{Bitbol16, Bitbol18, Gandarilla23, Lupo2024}.
\end{enumerate}

Together, these considerations motivated us to develop a bootstrapped approach to differentiable pairing, that we call DiffPaSS.
Briefly, after a single gradient step using the procedure outlined above and the initialization described in point \ref{item:first_grad_step} above, a number $n$ of pairings from the corresponding hard permutation is sampled uniformly at random, and used as fixed context in another run of gradient optimization using a single gradient step. The number
$n$ is gradually increased from $1$ to the size of the collections of sequences to be paired, with a step size $\Delta n$, whose default value is one.
\cref{fig:diffpass_video_screenshot} illustrates the first two iterations of this method.
See the Supplementary material for more details. Note that in this bootstrap process, it is possible that some incorrect pairs are selected to be used as fixed context. This can also happen in existing methods such as IPA~\cite{Bitbol16,Bitbol18}, where it was found not to be too detrimental in practice, partly because coevolutionary signal from correct pairs adds constructively, while noise from incorrect pairs does not~\cite{Gandarilla20}. 

\begin{figure}[htb]
    \centering
    \includegraphics[width=0.9\textwidth]{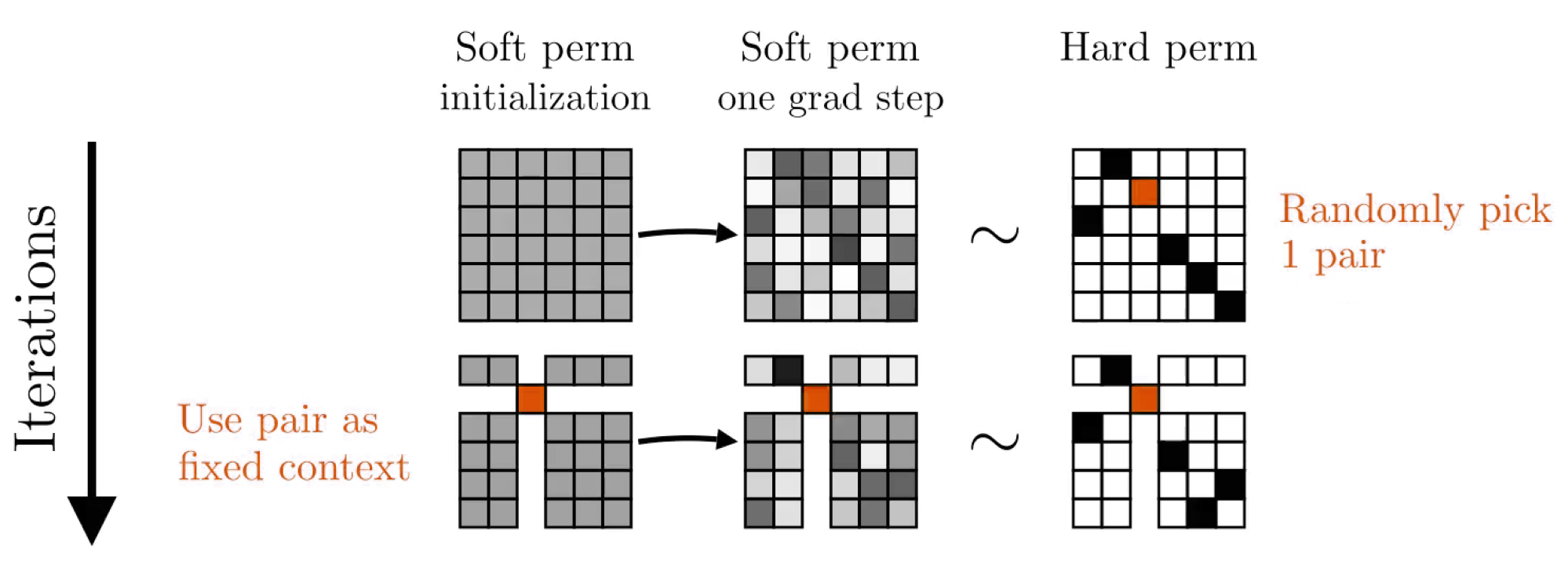}
    \caption{\textbf{First two iterations of a DiffPaSS bootstrap.} In this example, we pair sequences in a single species of size 6.
    Matrix entries are represented either as shades of gray -- ranging from white (when the entry is $0$) to black (when the entry is $1$) -- or in orange (also indicating a value of $1$) when they identify pairs randomly sampled at the end of one iteration and used as fixed pairs in the next.
    A tilde ($\sim$) between a soft and a hard permutation indicates that the soft permutation is ``close'' to the hard permutation, see \cref{eq:mena_limit}.
    An animation illustrating the full DiffPaSS bootstrap in this example is available at \url{https://www.youtube.com/watch?v=G2rV4ldgTIY}.}
    \label{fig:diffpass_video_screenshot}
\end{figure}

Using a single gradient step starting from our special initializations makes each step of the DiffPaSS bootstrap independent of gradient optimization hyperparameters such as learning rate and regularization strength.
It also makes it independent of the two hyperparameters needed to define the (truncated) Sinkhorn operator, see Supplementary material. Thus, in the DiffPaSS bootstrapped optimization process, the only parameter that can be tuned is the step size $\Delta n$.

\subsection*{Mutual-information--based scores for MSAs}
\label{msas_info_scores}

When $\mathcal{M}_\mathrm{A}$ and $\mathcal{M}_\mathrm{B}$ are MSAs, denoting by $\mathcal{M}_{\mathrm{A}, i}$ the $i$-th column of $\mathcal{M}_{\mathrm{A}}$ (and analogously for $\mathcal{M}_\mathrm{B}$), we define the \emph{inter-chain mutual information} score $\mathscr{S}_\mathrm{MI}(\mathcal{M}_\mathrm{A}, \mathcal{M}_\mathrm{B})$ by summing the mutual information estimates of all pairs of columns composed of one column in $\mathcal{M}_{\mathrm{A}}$ and one in $\mathcal{M}_{\mathrm{B}}$~\cite{Bitbol18}:
\begin{equation}
\label{eq:icmi}
    \mathscr{S}_\mathrm{MI}(\mathcal{M}_\mathrm{A}, \mathcal{M}_\mathrm{B}) = \sum_{i, j} \mathrm{I}(\mathcal{M}_{\mathrm{A}, i}; \mathcal{M}_{\mathrm{B}, j}) = \sum_{i, j} [\mathrm{H}(\mathcal{M}_{\mathrm{A}, i}) + \mathrm{H}(\mathcal{M}_{\mathrm{B}, j}) - \mathrm{H}(\mathcal{M}_{\mathrm{A}, i}, \mathcal{M}_{\mathrm{B}, j})]\,,
\end{equation}
where $\mathrm{I}(\cdot; \cdot)$ denotes the mutual information between two random variables, and $\mathrm{H}$ denotes the Shannon entropy.
In practice, we replace these information quantities by their plug-in estimates, i.e.\ we use observed frequencies instead of probabilities (see~\cite{Bitbol18}). 
Since, for any $i$ and permutation $\pi$, $\mathrm{H}(\mathcal{M}_{\mathrm{A}, i}) = \mathrm{H}(\pi(\mathcal{M}_\mathrm{A})_i)$, maximising $ \mathscr{S}_\mathrm{MI}(\pi)$ over permutations $\pi$ is equivalent to minimizing the \emph{inter-chain two-body entropy loss} $\mathscr{L}_\mathrm{2BE}(\pi)$ over permutations, with $\mathscr{L}_\mathrm{2BE}(\pi)$ defined as
\begin{equation}
\label{eq:ic2be}
    \mathscr{L}_\mathrm{2BE}(\pi) = \sum_{i, j} \mathrm{H}(\pi(\mathcal{M}_\mathrm{A})_i, \mathcal{M}_{\mathrm{B}, j}).
\end{equation}

We define a smooth extension $\hat{\mathscr{L}}_\mathrm{2BE}$ of $\mathscr{L}_\mathrm{2BE}$ to soft permutations as follows.
First, we represent the MSAs using one-hot encoding.
Namely, let $\bm{m}_{\mathrm{A}, i, n}$ denote the one-hot vector corresponding to row (i.e.\ sequence) $n$ and column (i.e.\ site) $i$ in $\mathcal{M}_\mathrm{A}$ -- and similarly for $\mathcal{M}_\mathrm{B}$.
The matrix of observed counts of all joint amino-acid states at the column pair $(i, j)$ can then be computed as $\sum_n \bm{m}_{\mathrm{A}, i, n} \otimes \bm{m}_{\mathrm{B}, j, n}$, where $\otimes$ denotes vector outer product.
As this expression is well-defined and smooth for pairs of arbitrary vectors, it yields a smooth extension of counts and frequencies provided that all vector entries are non-negative. This leads to a smooth extension $\hat{\mathrm{H}}(\cdot, \cdot)$ of the two-body entropy $\mathrm{H}(\cdot, \cdot)$.
In our case, for a soft permutation $\hat{\pi}$ represented as a matrix $\hat{P}$, we introduce a \textit{soft MSA} as $\hat{\pi}(\mathcal{M}_\mathrm{A}) = \hat{P} \bm{M}_\mathrm{A}$, where $\bm{M}_\mathrm{A}$ is the representation of $\mathcal{M}_\mathrm{A}$ as a tensor with an additional one-hot dimension.
We thus define the following smooth extension $\hat{\mathscr{L}}_\mathrm{2BE}$ of the inter-chain two-body entropy loss $\mathscr{L}_\mathrm{2BE}$:
\begin{equation}
\label{eq:ext_ic2be}
    \hat{\mathscr{L}}_\mathrm{2BE}(\hat{\pi}) = \sum_{i, j} \hat{\mathrm{H}}((\hat{P} \bm{M}_\mathrm{A})_i, \bm{M}_{\mathrm{B}, j})\,.
\end{equation}

\subsection*{Other scores}
\label{other_scores}

So far, we motivated the DiffPaSS bootstrap framework using the data structure of MSAs and the MI between MSA columns.
This was based on the observation \cite{Bitbol18} that MI contains useful signal for matching paralogs between interacting protein families.
However, a variety of other scores can also be used for paralog matching and for more general pairing problems, including scores based on sequence similarities, orthology, and phylogeny~\cite{goh2002co,Ramani03,Gertz03,Izarzugaza06,tillier2006codep,Izarzugaza08,Tillier09,Bradde10,Hajirasouliha12,ElKebir13}.
In some cases, these alternative scores are available even when alignments are not easy to construct or even meaningful.
Importantly, the DiffPaSS framework is a general approach that can be applied to different scores.
As an example, we extend the DiffPaSS framework to optimize graph alignment scores, and we consider applications to both aligned and non-aligned sequences.

\paragraph{Graph alignment scores.} 
Let us consider two ordered collections of sequences $\mathcal{M}_\mathrm{A}$ and $\mathcal{M}_\mathrm{B}$. Let us define a weighted graph $\mathcal{G}_A$ (resp.\ $\mathcal{G}_B$), whose nodes represent the sequences in $\mathcal{M}_\mathrm{A}$ (resp.\ $\mathcal{M}_\mathrm{B}$), and whose pairwise weights are stored in a matrix $\mathcal{W}_A$ (resp.\ $\mathcal{W}_B$).
Graph alignment (GA) can be performed between $\mathcal{G}_A$ and $\mathcal{G}_B$ using a variety of loss functions \citep{petric2019got,maretic2022wasserstein,maretic2022fgot,Gandarilla23}. When the GA loss function $\mathscr{L}_{\mathrm{GA}}(\mathcal{W}_\mathrm{A}, \mathcal{W}_\mathrm{B})$ is differentiable, we propose the following variant of DiffPaSS: define $\hat{\mathscr{L}}_{\mathrm{GA}}(\hat{\pi}) = \mathscr{L}_{\mathrm{GA}}(\hat{P} \mathcal{W}_\mathrm{A} \hat{P}^\mathrm{T}, \mathcal{W}_\mathrm{B})$, where $\hat{\pi}$ is a soft permutation encoded by the square matrix $\hat{P}$.
This definition is natural in the special case where $\hat{\pi}$ is a hard permutation, as $\hat{P} \mathcal{W}_\mathrm{A} \hat{P}^\mathrm{T}$ is then the matrix obtained from $\mathcal{W}_\mathrm{A}$ by permuting its rows and columns by $\hat{\pi}$.
Then, we perform a DiffPaSS bootstrap procedure (see above), using $\hat{\mathscr{L}}$ as the differentiable loss.

Here, as in Refs.~\cite{Bradde10,Gandarilla23}, we align $k$-nearest neighbor graphs. Specifically, we use pairwise weight matrices $\mathcal{W}$ constructed from ordered collections $\mathcal{M} = (s_i)_{i}$ of sequences as follows. Considering a symmetric distance (or dissimilarity) metric $d$ between sequences and an integer $k > 1$, we define the $(i, j)$-th entry of $\mathcal{W}$ as
\begin{equation}
    \label{eq:knn_weight_mat}
    \mathcal{W}_{ij} = \begin{cases} e^{-d(s_i, s_j) / D^2} & \text{if $s_i$ is among the $k$ nearest neighbors of $s_j$, or vice-versa,} \\ 0 & \text{otherwise.} \end{cases}
\end{equation}
Here, $D$ is the average distance (over all sequences in $\mathcal{M}$) of the $k$-th nearest neighbor sequence.
As distance metric, we use Hamming distances if the sequences are aligned, and edit distances otherwise.
We set $k = 20$ or $k = 30$.
Finally, as in Refs.~\cite{Bradde10,Gandarilla23}, we use
\begin{equation}
    \label{eq:GA_loss}
    \mathscr{L}_{\mathrm{GA}}(\mathcal{W}_\mathrm{A}, \mathcal{W}_\mathrm{B}) = -\sum_{i} \sum_{j > i} (\mathcal{W}_\mathrm{A})_{ij} (\mathcal{W}_\mathrm{B})_{ij}
\end{equation}
as a GA loss function.

\subsection*{Robust pairs and iterative variants of DiffPaSS}
\label{robust_diffpass-ipa}

We empirically found that, when running a DiffPaSS bootstrap, some sequence pairs are found by all the hard permutations explored.
We call these \emph{robust pairs}, and notice that they tend to have high precision.
This is illustrated in \cref{fig:precision_recall} for a benchmark prokaryotic dataset described in Supplementary material Section \ref{sec:Datasets}.
This suggests that they can be used as a starting set $\mathcal{F}_\mathrm{AB}$ of fixed pairs in another run of DiffPaSS, where further fixed pairs are (randomly) chosen from the remaining sequences, see above.
This process can be repeated several times by adding new robust pairs, found among the remaining sequences, to the set $\mathcal{F}_\mathrm{AB}$.
We call this iterative procedure DiffPaSS-IPA (``Iterative Pairing Algorithm'', following \citet{Bitbol16, Bitbol18}).
The final output of DiffPaSS-IPA is the hard permutation with lowest observed loss across all IPA runs.
In practice, for all IPA variants of DiffPaSS, we use $N_\mathrm{IPA} = 3$ iterations here.

\section*{Results}
\label{results}

\subsection*{DiffPaSS accurately and efficiently pairs paralogs using MI}

\begin{figure}[htb]
    \centering
    \includegraphics[width=0.95\textwidth]{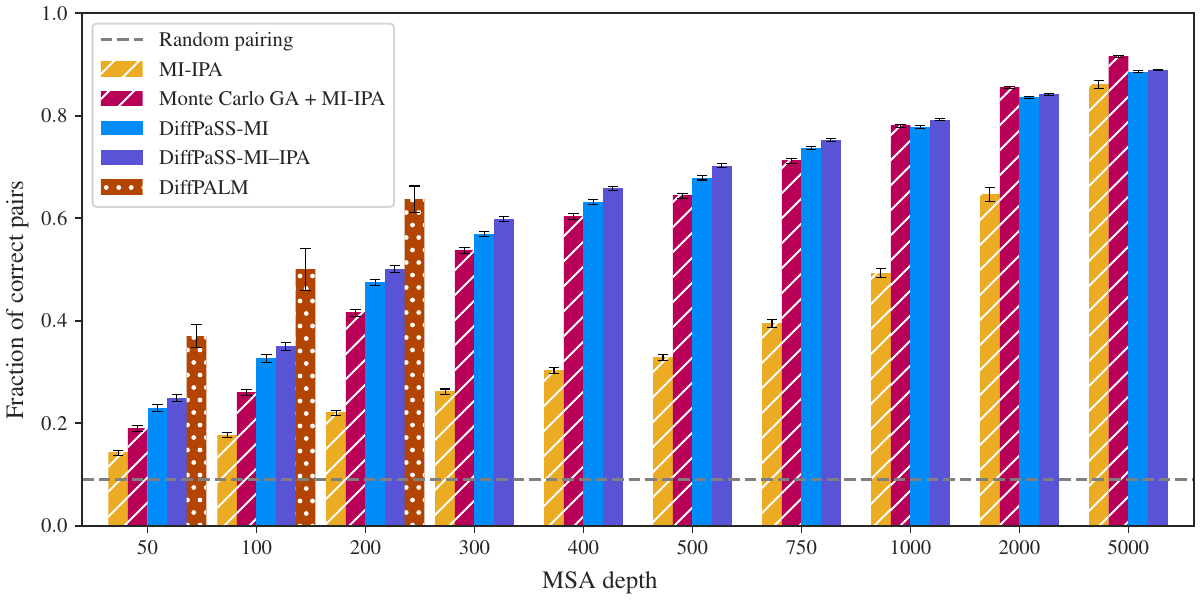}
    \caption{\textbf{Performance of coevolution-based pairing methods on HK-RR MSAs.} The performance of DiffPaSS-MI and DiffPaSS-MI-IPA is compared to that of MI-IPA~\cite{Bitbol18}, alone or combined with Monte Carlo GA~\cite{Gandarilla23}, and to that of DiffPALM~\cite{Lupo2024}, for HK-RR MSAs with different depths. The HK-RR dataset is described in Supplementary material Section \ref{sec:Datasets}. With all methods, a full one-to-one within-species pairing is produced, and performance is measured as the fraction of correct pairs among all predicted pairs.
    For the combined Monte Carlo GA and MI-IPA method, we used the same settings as \citet{Gandarilla23}. 
    Results for DiffPALM are taken from \citet{Lupo2024}. }
    \label{fig:hk-rr_results_comparison}
\end{figure}

\paragraph{DiffPaSS-MI pairs interacting sequences more accurately than other MI-based methods.}
Let us first consider coevolution-based pairing of partners from the MSAs of two interacting protein families. How does our differentiable pairing method compare to existing discrete pairing methods?
To address this question, we test DiffPaSS and DiffPaSS-IPA, using the total inter-chain MI as a score (see Methods), on a benchmark dataset composed of ubiquitous prokaryotic proteins from two-component signaling systems~\cite{Barakat09,Barakat11}, which enable bacteria to sense and respond to environment signals. This dataset comprises cognate pairs of histidine kinases (HKs) and response regulators (RRs) determined using genome proximity, see Supplementary material Section \ref{sec:Datasets}. These proteins feature high specificity and generally interact one-to-one~\cite{Laub07,Cheng14}. We blind all partnerships and we compare different coevolution-based methods on the task of pairing these interacting sequences within each species, without giving any known paired sequences as input.
\cref{fig:hk-rr_results_comparison} shows that both DiffPaSS and DiffPaSS-IPA significantly outperform the MI-IPA algorithm \cite{Bitbol18}, which performs a discrete approximate maximization of the same MI score. The gain of performance obtained by using DiffPaSS instead of MI-IPA is particularly remarkable for relatively shallow MSAs, up to $\sim $2000 sequences deep. Recall that MI-IPA is quite data-thirsty~\cite{Bitbol18}. 
Furthermore, \cref{fig:hk-rr_results_comparison} shows that, for MSAs up to depth 750 (resp.\ 1000), DiffPaSS (resp.\ DiffPaSS-IPA) outperforms a recent method which combines Monte-Carlo GA and MI-IPA \citep{Gandarilla23}.
This is significant because this combined GA and MI-IPA method exploits both phylogenetic sequence similarity via GA and coevolution via MI, while here we only used MI in DiffPaSS. 
However, the combination of GA and MI-IPA slightly outperforms MI maximization (with DiffPaSS) for deeper MSAs (depths 2000 and 5000) -- see ``DiffPaSS-MI is highly effective at extracting MI signal" for additional information. \cref{fig:ga_plus_diffpass} shows that, for deep alignments, the performance of DiffPaSS-MI is improved by combining it with GA, and becomes comparable with that of Monte Carlo GA combined with MI-IPA. Note that for shallow alignments, combining with GA improves the performance of MI-IPA, but not that of DiffPaSS-MI. 

An important parameter of GA is the number $k$ of nearest neighbors considered in the $k$ nearest-neighbor ($k$NN) graphs that are aligned, see \cref{eq:knn_weight_mat}. While in \citet{Gandarilla23} it was optimized on the same HK-RR dataset as the one we use here, this was for MSAs of depth 5000. In \cref{fig:gaipa_knn_test}, we investigate the impact of $k$ on the performance of MI-IPA combined with Monte Carlo GA for HK-RR MSAs of depth 100. We observe that smaller values of $k$ lead to slightly better performances for these shallower MSAs, but that the performance of DiffPaSS-MI--IPA remains higher than that of MI-IPA combined with Monte Carlo GA for all values of $k$ considered. 
Note that for deeper MSAs (depth 5000), \cref{fig:hk-rr_results_comparison} shows that all methods, including MI-IPA, perform well and obtain correct results for more than 80\% of the pairs~\cite{Bitbol18,Gandarilla23}.

While DiffPaSS-MI and MI-IPA are both based on MI, we recently introduced DiffPALM \citep{Lupo2024}, an approach that leverages the MSA-based protein language model MSA Transformer \citep{rao2021msa}. \cref{fig:hk-rr_results_comparison} shows that DiffPaSS does not reach the performance achieved by DiffPALM. However, DiffPaSS is several orders of magnitude faster (see below), and easily scales to much deeper MSA depths for which DiffPALM cannot be run due to memory limitations.

An attractive feature of the DiffPaSS bootstrap process is that the only hyperparameter that can be varied is the integer step size $\Delta n$ controlling the number of fixed pairs added at each bootstrap step (see Methods). By default, we choose a step size $\Delta n=1$. We found that it led to better performance of DiffPaSS-MI--IPA than using a step size $\Delta n=2$, see \cref{fig:step_size_2}. Note however that increasing $\Delta n$ can lead to significant speed-ups.

\paragraph{Extension to another benchmark prokaryotic dataset.} So far, we tested DiffPaSS on the HK-RR benchmark. In \cref{fig:malg-malk}, we show the performance of DiffPaSS-MI--IPA on a dataset of ABC transporter proteins homologous to the \textit{Escherichia coli} MALG-MALK pair of interacting proteins, see Refs.~\cite{Bitbol16,Bitbol18,Gandarilla23,Lupo2024}, for MSAs of depth 100. We find that in this case, the performance of DiffPaSS-MI-IPA is comparable to that of MI-IPA combined with Monte Carlo GA for various values of $k$, and to that of DiffPALM. This confirms that the ability of DiffPaSS at predicting interacting pairs generalizes beyond the HK-RR case. 

\begin{figure}[htbp]
    \centering
    \includegraphics[width=\textwidth]{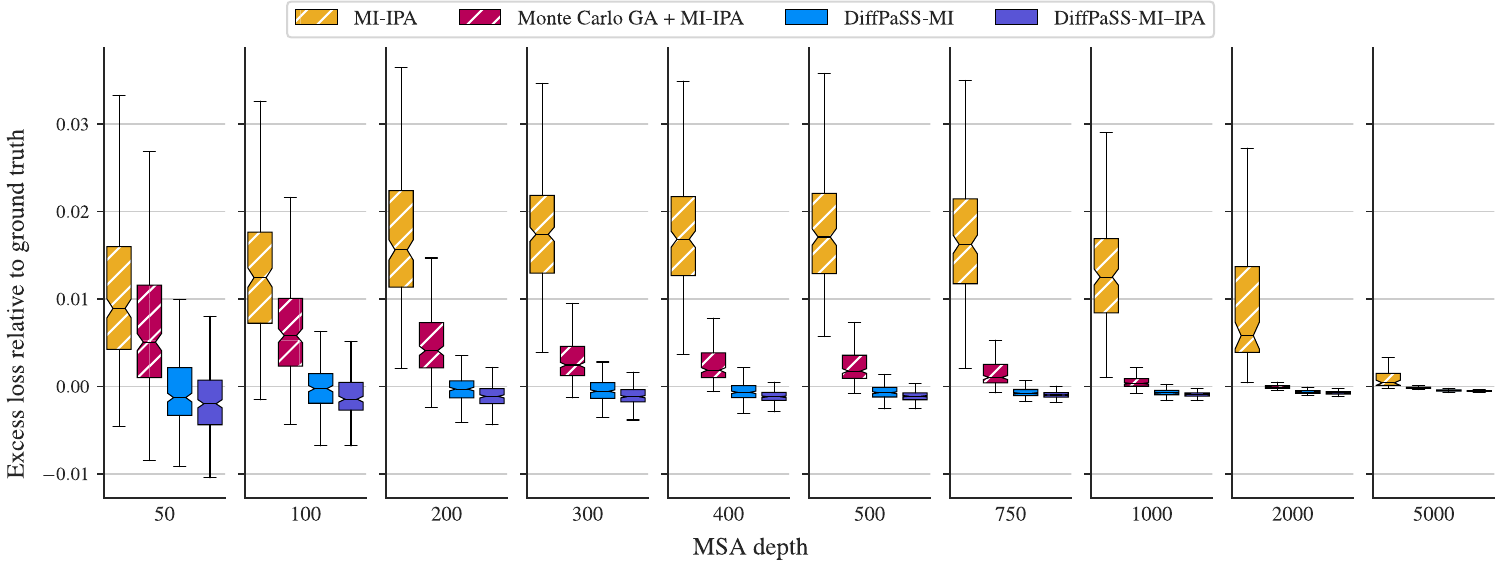}
    \caption{\textbf{Inter-chain two-body entropy losses for predicted pairings.} The distributions of the excess inter-chain two-body entropy losses of predicted pairings relative to ground-truth pairings are shown as box plots for several methods and several MSA depths, on the HK-RR dataset (same MSAs as in \cref{fig:hk-rr_results_comparison}).}
    \label{fig:hk-rr_excess_2be_comparison}
\end{figure}

\paragraph{DiffPaSS-MI is highly effective at extracting MI signal.} How does DiffPaSS achieve better performance than MI-IPA? It was shown in \citet{Bitbol18} that, for relatively shallow MSAs, the approximate discrete optimization algorithm used to maximize inter-chain MI (\cref{eq:icmi}) in the MI-IPA approach does not provide pairings that reach MI values as high as the ground-truth pairings.
Thus motivated, we compare the inter-chain MI of the paired MSAs produced by DiffPaSS, and by other methods, to the MI of the ground-truth pairings.
We use the difference between the inter-chain two-body entropy loss (\cref{eq:ic2be}) of a paired MSA and that of the ground-truth paired MSA. Indeed, this ``excess loss'' is equal to minus the difference between the corresponding inter-chain MIs, see \cref{eq:icmi,eq:ic2be}.
\cref{fig:hk-rr_excess_2be_comparison} shows the distribution of excess inter-chain two-body entropy losses for each method.
We observe that the median MI of DiffPaSS(-IPA) final pairings is indistinguishable or higher than that of the ground-truth pairings for all MSA depths considered (as excess losses are positive). Furthermore, the distribution of excess losses across MSAs becomes very narrow for deeper MSAs, showing that a score very close to the ground-truth MI is then systematically obtained. Meanwhile, for shallow MSAs, MI-IPA and the combined Monte Carlo GA + MI-IPA methods yield paired MSAs that have a substantially lower MI than the ground-truth pairings.
This improves for deeper MSAs, faster for the combined method than for MI-IPA (see also~\cite{Bitbol18} for MI-IPA).
Note however that, for all MSA depths, the combined method yields paired MSAs with higher excess losses than DiffPaSS-MI(--IPA), even though it produces a higher fraction of correct pairs, on average, for the deepest MSAs (depths 2000 and 5000, see \cref{fig:hk-rr_results_comparison}). 
This shows that, at least on this dataset, the \color{blue}inter-chain two-body entropy loss \color{black}has some limitations for the pairing of interaction partners, \color{blue}in the sense that the correctly paired MSA has a somewhat higher loss (i.e., lower MI) than some slightly different ones. This might arise due to phylogenetic relationships between sequences. Indeed, \color{black}exploiting sequence similarity and phylogeny via GA~\cite{Gandarilla23} provides additional information. 
Recall that \cref{fig:ga_plus_diffpass} shows that this information can be exploited when using DiffPaSS-MI as well as when using MI-IPA.

While our focus is here on comparing to the ground-truth pairings, the distributions of inter-chain two-body losses are shown in \cref{fig:hk-rr_2be_comparison}, for all methods as well as for random pairings.
This figure shows the amount of available MI signal for pairing in these datasets.
Overall, these results demonstrate that DiffPaSS is extremely effective at extracting the available MI signal.

\paragraph{DiffPaSS-MI identifies robust pairs.}The combined method Monte Carlo GA + MI-IPA relies on identifying robust pairs which are predicted by multiple independent runs of GA and then using them as a training set of known pairs for MI-IPA.
\cref{fig:precision_recall} compares these robust pairs with those that are identified by DiffPaSS-MI(-IPA), specifically the set $\mathcal{F}_{\mathrm{AB}}$ after 3 IPA iterations, see Methods. It shows that, for relatively shallow MSAs, DiffPaSS-IPA typically identifies fewer correct pairs as robust than Monte Carlo GA, but its robust pairs are correct more often. For deeper MSAs, the number of correct robust pairs identified by DiffPaSS-MI--IPA sharply increases, while it plateaus for Monte Carlo GA.

\paragraph{DiffPaSS-MI is substantially faster than other approaches.} How does DiffPaSS compare to existing algorithms in terms of computational runtime? DiffPaSS can easily be run on a modern GPU, while MI-IPA and Monte Carlo GA + MI-IPA are CPU-only algorithms.
\cref{fig:hk-rr_diffpass_ga-ipa_runtime_comparison} demonstrates that DiffPaSS-MI is considerably faster than Monte Carlo GA + MI-IPA across all the MSA depths we analyzed.
DiffPALM \citep{Lupo2024} (not shown in \cref{fig:hk-rr_diffpass_ga-ipa_runtime_comparison}) has much longer runtimes than both methods, taking e.g.\ over three orders of magnitudes longer than DiffPaSS to pair HK-RR MSAs of depth $\sim 50$. Thus, a considerable asset of DiffPaSS is its rapidity, which makes it scalable to large datasets comprising many pairs of protein families.

\subsection*{DiffPaSS improves the structure prediction by AlphaFold-Multimer of some eukaryotic complexes}
\label{afm_results}
\begin{figure}[htbp]
    \centering
    \includegraphics[width=0.95\textwidth]{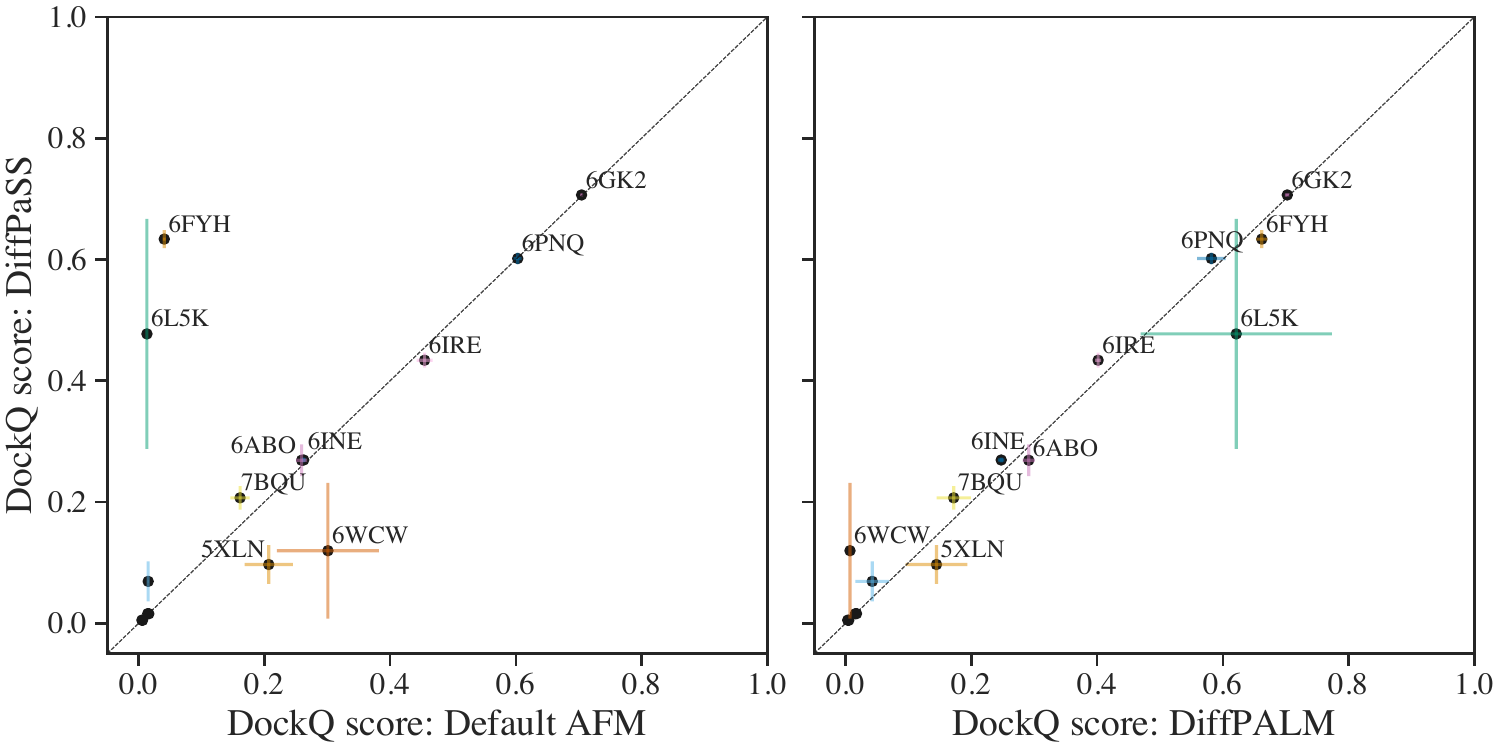}
    \caption{\textbf{Performance of structure prediction by AFM using different MSA pairing methods.} We report the performance of AFM, in terms of DockQ scores, for the 15 complexes considered in \cite{Lupo2024}, using different pairing methods on the same MSAs. Left panel: DiffPaSS versus default AFM pairing. Right panel: DiffPaSS versus DiffPALM. For each complex, AFM is run five times, and the top predicted structure by AFM confidence is considered each time, yielding 5 predicted structures in total. DockQ scores are averaged over the 5 predictions and standard errors are shown as error bars. Points with DockQ below $0.1$ are not labelled with their PDB ID for graphical reasons. 
    }
    \label{fig:dockq_comparison_simple}
\end{figure}

While genome proximity can often be used to pair interaction partners in prokaryotes, it is not the case in eukaryotes. Pairing correct interaction partners is thus a challenging problem in eukaryotes, which also often have many paralogs per species \cite{Makarova05} while eukaryotic-specific protein families generally have fewer total homologs and smaller diversity than in prokaryotes. Solving this problem has an important application to the prediction of protein complex structure. Indeed, AlphaFold-Multimer (AFM) \cite{evans2021protein} relies on paired MSAs~\cite{evans2021protein,bryant2022improved}.
Can DiffPaSS improve complex structure prediction by AFM \citep{evans2021protein} in eukaryotic complexes? 
As a first exploration of this question, we consider the same 15 eukaryotic complexes as in \citet{Lupo2024}, where improvements over the default AFM pairing methods were reported by pairing using the MSA-Transformer--based method DiffPALM \citep{rao2021msa}.
More information on these structures and on the AFM setup can be found in Supplementary material Sections \ref{supp_datasets_eukaryotic} and~\ref{sec:generalities_AFM}.
\cref{fig:dockq_comparison_simple} compares the performance of AFM on these complexes, using three different pairing methods (default AFM, DiffPALM, and DiffPaSS-MI) on the same initial unpaired MSAs. 
We use the DockQ score, a widely used measure of quality for protein-protein docking \citep{Basu16}, as a performance metric for complex structure prediction.
More details are given in \cref{fig:dockq_comparison}, where we include the top 5 predicted structures of each run of AFM. 
AFM confidence scores for these predictions are also shown in \cref{fig:afm_confidences}.
These results show that DiffPaSS can improve complex structure prediction in some cases. Furthermore, our results are consistent with those we previously obtained with DiffPALM~\cite{Lupo2024}: the two structures that are substantially improved are 6FYH and 6L5K. 

\subsection*{DiffPaSS allows accurate graph alignment}

\begin{figure}[htb]
    \centering
    \includegraphics[width=0.95\textwidth]{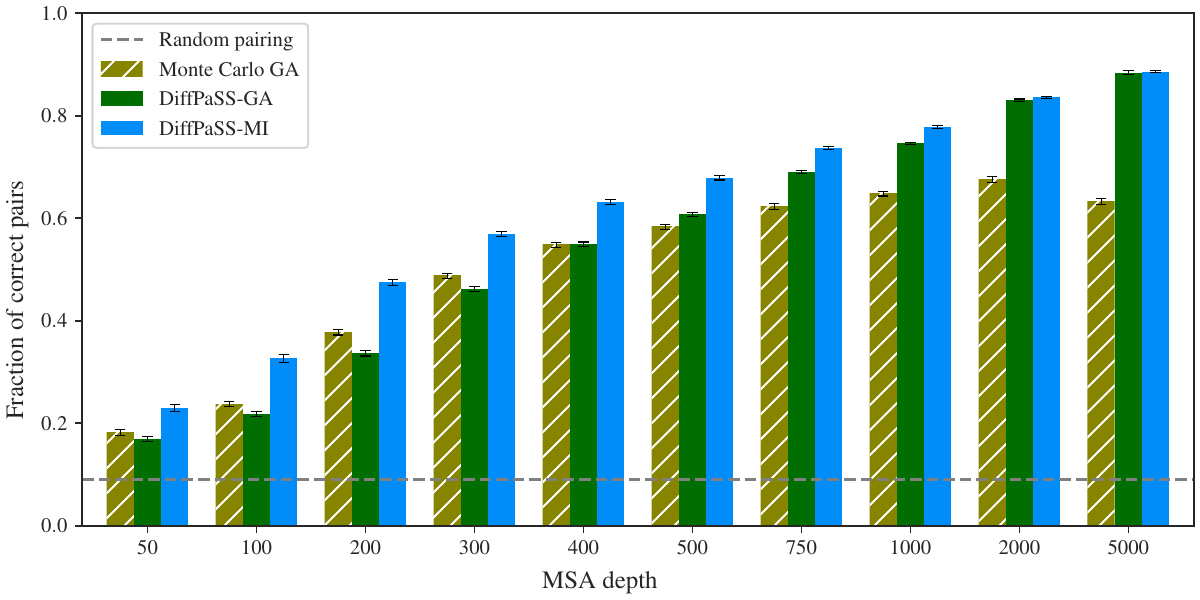}
    \caption{\textbf{Performance of GA- or MI-based pairing methods on HK-RR MSAs.} The fraction of correct pairs obtained is shown for DiffPaSS-GA and for the Monte Carlo-based GA implementation from~\cite{Gandarilla23}, on the same HK-RR MSAs as in \cref{fig:hk-rr_results_comparison}.
    The results for DiffPaSS-MI from \cref{fig:hk-rr_results_comparison} are shown for comparison.
    }
    \label{fig:hk-rr_ga_results_comparison}
\end{figure}

So far, we used DiffPaSS for the specific problem of paralog matching from the MSAs of two interacting protein families, using coevolution measured via MI. However, the optimization approach of DiffPaSS is general and can be applied to other scores, and to non-aligned sequences. In particular, it can be used for the problem of graph alignment, see Methods. To assess how well DiffPaSS performs at this problem, we first return to our benchmark HK-RR dataset.
Instead of the inter-chain MI, to predict pairs we now use the GA loss~\cite{Bradde10,Gandarilla23}, defined in \cref{eq:GA_loss}. Note that we use Hamming distances to define the weight matrices $\mathcal{W}_\mathrm{A}$ and $\mathcal{W}_\mathrm{B}$. Fig.~\ref{fig:hk-rr_ga_results_comparison} shows that DiffPaSS-GA is competitive with the Monte Carlo simulated annealing algorithm in \citet{Gandarilla23}, which optimizes the same GA loss. More precisely, DiffPaSS-GA performs slightly less well than Monte Carlo GA for shallow MSAs, but better than it for deeper ones, where the GA problem becomes trickier as the neighbor graph contains more and more inter-species edges. We also compare DiffPaSS-GA with DiffPaSS-MI, and find that DiffPaSS-MI outperforms DiffPaSS-GA for shallow MSAs, but performance becomes similar for deeper ones.

Contrary to MI, GA does not require sequences to be aligned. Indeed, one can use a distance measure between sequences which does not involve aligning them. This opens the way to broader applications. An interesting one, which is close to the paralog matching problem studied here, regards pairing T cell receptor chains. Specifically, we consider collections of sequences from hypervariable CDR3$\alpha$ and CDR3$\beta$ loops in TCRs binding to a fixed epitope, as in \citet{milighetti2024intra}. The task is to pair each $\alpha$ chain to its $\beta$ chain within each human patient. Thus, here, patients play the part of species in our paralog matching problem. Our dataset of paired CDR3$\alpha$-CDR3$\beta$ loop sequences from TCRs is described in Supplementary material Section \ref{sec:Datasets}. These collections of CDR3$\alpha$ and CDR3$\beta$ sequences are difficult to align due to their hypervariability, motivating the choice of GA for this problem, using the GA loss in \cref{eq:GA_loss} with weight matrices $\mathcal{W}_\mathrm{A}$ and $\mathcal{W}_\mathrm{B}$ defined using edit distances instead of Hamming distances (see \cref{eq:knn_weight_mat}).
Note also that here we use $k = 20$ nearest neighbors, in order to compare our results with those from \citet{milighetti2024intra}, obtained using the same value of $k$.
The differences between the losses obtained by DiffPaSS and those obtained by the Monte Carlo GA algorithm are shown in \cref{fig:CDR_diffpass_vs_simulated_annealing}.
DiffPaSS achieves substantially lower losses than the Monte Carlo GA algorithm, on average, in the four datasets containing the largest numbers of sequences to pair (from $\sim$700 to $\sim$2000).
On the other hand, it yields slightly higher losses than Monte Carlo GA, on average, for datasets with 500 pairs or fewer.
The good performance of DiffPaSS-GA on larger datasets is consistent with our HK-RR results (\cref{fig:hk-rr_ga_results_comparison}).

\newpage
\section*{Discussion}
\label{discussion}

We introduced DiffPaSS, a framework to pair interacting partners among two collections of biological sequences. DiffPaSS uses a versatile and hyperparameter-free differentiable optimization method that can be applied to various scores. It outperforms existing discrete optimization methods for pairing paralogs using MI, most spectacularly in the regime of shallow alignments. Strikingly, on our benchmark datasets, DiffPaSS is able to extract all the MI signal available for pairing. DiffPaSS is computationally highly efficient, compared to existing discrete optimization methods and, by far, to DiffPALM \cite{Lupo2024}, our pairing method based on MSA Transformer. Thus, it has the potential be applied to large datasets. Our first explorations on a small set of eukaryotic  protein complexes show that paired alignments produced by DiffPaSS may lead to improvements of complex structure prediction by AlphaFold-Multimer \cite{evans2021protein}. 
DiffPaSS does not need to start from aligned sequences, thus opening the way to broader applications. Our results on pairing TCR chains show promise for the use of DiffPaSS to optimize scores that cannot be constructed from MSAs.

Since DiffPaSS shows promise for structural biology, it would be very interesting to apply it more broadly to protein complex structure prediction problems. Besides, applications to TCRs are also promising. We expect CDR3$\alpha$-CDR3$\beta$ paired sequences to show lower co-evolution signal than other interacting protein families. Indeed, they are generated at random and selected for their joint ability to bind a given antigen, rather than directly co-evolving. Development of similarity measures that can cluster TCRs with the same specificity in sequence space is ongoing~\cite{Thomas14,Dash17,Thakkar19,Vujovic20,Wu21,Meysman23,Nagano24}. Using DiffPaSS-GA with these similarity measures could allow testing of metrics specifically optimised to capture TCR co-specificity. 

One limitation of DiffPaSS in its current formulation is that it assumes one-to-one pairings.
It would be very interesting to extend it to pairing problems between collections of different sizes.
To achieve this, one could e.g.\ replace the Sinkhorn operator with Dykstra's operator, see \citet{maretic2022wasserstein}.

Finally, despite DiffPaSS's ability to extract all available MI signal from our benchmark dataset, we found that our MSA-Transformer--based pairing method DiffPALM outperforms DiffPaSS. This shows that MSA-based protein language models such as MSA Transformer truly capture more signal usable for pairing than MI.

\section*{Acknowledgments}
U.~L., D.~S., and A.-F.~B.\ acknowledge funding from the European Research Council (ERC) under the European Union’s Horizon 2020 research and innovation programme (grant agreement No.~851173, to A.-F.~B.).
M.~M.\ acknowledges funding by Cancer Research UK through a Non-Clinical Training Award [A29287].

\newpage
\appendix

\begin{center}
\LARGE{\textbf{Supplementary material}}
\end{center}
\vspace{0.2cm}

\renewcommand{\thesection}{\arabic{section}}
\renewcommand{\thefigure}{S\arabic{figure}}
\setcounter{figure}{0}
\renewcommand{\thetable}{S\arabic{table}}
\setcounter{table}{0}
\renewcommand{\theequation}{S\arabic{equation}}
\setcounter{equation}{0}

\tableofcontents

\section{Detailed methods}
\label{appx_methods}

In this section, we present a more detailed account of the methods developed in this work. We acknowledge that there are some redundancies with the main text methods. We retained them for the sake of completeness and readability of this Supplementary material.

\subsection{Preliminaries}
\label{preliminaries}

Let $\mathcal{M}_\mathrm{A}$ and $\mathcal{M}_\mathrm{B}$ be ordered collections of amino-acid sequences that are partitioned into $K$ groups, each of size $N_k$ where $k = 1, \ldots, K$.
In the important special case where $\mathcal{M}_\mathrm{A}$ and $\mathcal{M}_\mathrm{B}$ are collections of proteins from two interacting protein families, the ``groups'' will be species, with species $k$ assumed to contain $N_k$ paralogous proteins in both families.

Let $\mathscr{S}$ be a score function of the two ordered collections.
We would like to find a permutation $\pi$ of the entries in $\mathcal{M}_\mathrm{A}$ which maximises $\mathscr{S}(\pi(\mathcal{M}_\mathrm{A}), \mathcal{M}_\mathrm{B})$, under the constraint that $\pi$ does not send sequences from any one group into a different group.
A priori, there are $\prod_{k=1}^{K} N_k !$ permutations satisfying this constraint.
Note that we can equivalently describe this as the problem of finding $\mathscr{S}$-maximising one-to-one matchings between the sequences in $\mathcal{M}_\mathrm{A}$ and those in $\mathcal{M}_\mathrm{B}$.
Since $\mathcal{M}_\mathrm{A}$ and $\mathcal{M}_\mathrm{B}$ will remain fixed, by abusing notation we denote $\mathscr{S}(\pi(\mathcal{M}_\mathrm{A}), \mathcal{M}_\mathrm{B})$ simply by $\mathscr{S}(\pi)$. 

Let $N > 0$, and denote the set of permutation matrices of $N$ elements by $\mathcal{P}_N$.
For any integer $\ell > 0$ and arbitrary $N \times N$ square matrix $X$, define the \emph{$\ell$-truncated Sinkhorn operator}
\begin{equation}
\label{eq:sinkhorn_op}
    S^{\ell}(X) = (\mathcal{C} \circ \mathcal{R})^\ell(\exp(X))
\end{equation}
consisting of first applying the componentwise exponential function $\exp$ to $X$, and then iteratively normalizing rows ($\mathcal{R}$) and columns ($\mathcal{C}$) $\ell$ times.
It can be shown \citep{Mena2018} that $\displaystyle S(X) := \lim_{\ell \to \infty} S^{\ell}(X)$ defines a smooth operator mapping to bistochastic matrices\footnote{A bistochastic matrix is a matrix with non-negative entries whose all rows and columns sum to $1$.} and that, for almost all $X$,
\begin{equation}
    \label{eq:mena_limit}
    \lim_{\tau \to 0^+} S(X / \tau) = M(X) := \textrm{argmax}_{P \in \mathcal{P}_N} [\mathrm{trace}(P^{\mathrm{T}} X)].
\end{equation}
The operator $M$ defined in \cref{eq:mena_limit}, which maps onto permutation matrices, can be computed using standard discrete algorithms for linear assignment problems \citep{Kuhn55}.
\cref{eq:mena_limit} implies that, using the ``parameterization matrices'' $X$, one can smoothly navigate the space $\mathcal{B}_N$ of all $N \times N$ bistochastic (resp.\ near-bistochastic) matrices using $S$ (resp.\ $S^\ell$), while keeping track of the ``nearest'' permutation matrices using $M$.
In what follows, we will refer to (near-)bistochastic matrices as ``soft permutations'' and to true permutations as ``hard permutations''. 

We may therefore hope to find optimal hard permutations for the original score $\mathscr{S}$ by optimizing a suitable smooth extension $\hat{\mathscr{S}}$ of $\mathscr{S}$ to soft permutations, since this can be done efficiently using gradient methods. 
In general, optimization of $\hat{\mathscr{S}}$ can be very sensitive to hyperparameters such as the ``temperature'' $\tau > 0$ in \cref{eq:mena_limit}, the standard deviation of the entries of $X$ at initialization, the optimizer learning rate, and the strength of regularization.

\subsection{Initialization and bootstrapped optimization}
\label{initialization_bootstrap}

We found that solving $X^* = \textrm{argmin}_X \hat{\mathscr{L}}_\mathrm{2BE}(S^{\ell}(X / \tau))$, for some choice of $\ell$ and $\tau$, using several steps of gradient descent, generally yields sub-optimal hard permutations $M(X^*)$ for the original loss $\mathscr{L}_\mathrm{2BE}$, see \cref{eq:mena_limit}.
Informally, this is because hard permutations are local minima for $\hat{\mathscr{L}}$, a situation reminiscent of how the entropy of a single Bernoulli random variable with parameter $p$ has minima at the ``sharp'' cases $p = 0$ and $p = 1$.
Nevertheless, we found that outcomes improved when all entries of $X$ are initialized to be zero.
Indeed, as shown in \cref{fig:hk-rr_precision_first_gradient_step} for the benchmark HK-RR prokaryotic dataset described in Supplementary material Section \ref{sec:Datasets}, the \emph{first} gradient step alone, when $X \equiv 0$ at initialization, is often competitive with a full-blown discrete algorithm for approximate maximization of $\mathscr{S}_\mathrm{MI}$, called MI-IPA \citep{Bitbol18}.
Finally, as previously observed using several methods \citep{Bitbol16, Bitbol18, Gandarilla23, Lupo2024}, pairing performance can be expected to increase if correct pairings are used as fixed context, biasing the computation of the two-body entropies.
Together, these considerations led us to DiffPaSS, our proposed bootstrapped approach to differentiable pairing.
Given a loss $\mathscr{L}$ and its smooth extension $\hat{\mathscr{L}}$, two ordered collections $\mathcal{M}_\mathrm{A}$ and $\mathcal{M}_\mathrm{B}$ containing $D$ sequences, and (optionally) a pre-existing set $\mathcal{F}_\mathrm{AB} = \{ (a_i, b_i) \}_{i=1}^{D_\mathrm{fix}}$ containing $D_\mathrm{fix}$ matched pairs of sequences, we proceed as follows.\footnote{An animation illustrating this algorithm in the special case $D_{\mathrm{fix}} = 0$ is available at the following URL: \url{https://www.youtube.com/watch?v=G2rV4ldgTIY}.}
Let $\Delta n$ be a positive integer, referred to in the main text as the step size, and $N_\mathrm{steps} \geq 1$ be such that $(N_\mathrm{steps} - 1)\Delta n$ is the largest multiple of $\Delta n$ that is less than $D - D_\mathrm{fix}$. Initialize $\mathcal{F}' = \mathcal{F}_\mathrm{AB}$ and $n_\mathrm{prev} = 0$; then, for every 
$n = \Delta n,\,2\Delta n,\ldots, (N_\mathrm{steps} - 1) \Delta n, D - D_\mathrm{fix}$:
\begin{enumerate}
     \item if $n_\mathrm{prev} = D - D_\mathrm{fix}$, terminate;
    \item define $P'$ as the $(D_\mathrm{fix} + n_\mathrm{prev}) \times (D_\mathrm{fix} + n_\mathrm{prev})$ permutation matrix corresponding to the matchings in $\mathcal{F}'$;
    \item initialize two zero $D \times D$ matrices $P$ and $\hat{P}$, and copy $P'$ into the row-column pairs belonging to $\mathcal{F}'$ in both cases;
    \item initialize a $(D - D_\mathrm{fix} - n_\mathrm{prev}) \times (D - D_\mathrm{fix} - n_\mathrm{prev})$ parameterization matrix $X \equiv 0$, to be used for sequences not involved in pairs in $\mathcal{F}'$; 
    \item update $X \leftarrow - \nabla (\hat{\mathscr{L}} \circ \tilde{S^\ell})(X = 0)$, where $\tilde{S^\ell}(X)$ denotes copying $S^\ell(X)$ into the submatrix of $\hat{P}$ obtained by removing rows and columns involved in pairs in $\mathcal{F}'$;
    \item\label{step:hard_perm} compute the hard permutation $\pi = \tilde{M}(X)$, where $ \tilde{M}(X)$ denotes copying $M(X)$ into the submatrix of $P$ obtained by removing rows and columns involved in pairs in $\mathcal{F}'$;
    \item\label{step:fixed_pairs} pick $n$ pairs of sequences matched by $\pi$, but not in $\mathcal{F}_\mathrm{AB}$, uniformly at random;
    \item update $\mathcal{F}' \leftarrow \mathcal{F}_\mathrm{AB} \cup \{ \text{pairs selected in step \ref{step:fixed_pairs}} \}$;
    \item update $n_\mathrm{prev} \leftarrow n$.
\end{enumerate}
For every $n = \Delta n,\,2\Delta n,\ldots, (N_\mathrm{steps} - 1) \Delta n, D - D_\mathrm{fix}$, we record the loss $\mathscr{L}(\pi)$ at step \ref{step:hard_perm}, and the final output of DiffPaSS is the hard permutation $\pi^*$ corresponding to the lowest recorded loss.

Note that the gradient of $S^{\ell}(X / \tau)$, evaluated at $X = 0$, only changes by a global scale factor if $\tau$ is changed, and the matching operator $M$ is scale-invariant.
Hence, we can set $\tau = 1$ as the obtained hard permutations are independent of it.
Similarly, all hard permutations obtained are independent of the choice of learning rate and regularization strength.
Perhaps more surprisingly, for any $\ell > 1$, the gradient of $S^{\ell}$ evaluated at $X = 0$ is equal to corresponding gradient of $S^{\ell = 1}$.
We prove this in Supplementary material Section \ref{sec:Theorem_Sinkhorn_normalizations}.
Hence, we can set $\ell = 1$ throughout, leading to significant runtime gains.

\paragraph{Robust pairs and DiffPaSS-IPA.}
\label{appx_robust_diffpass-ipa}

\begin{figure}[h!]
    \centering
    \includegraphics[width=\textwidth]{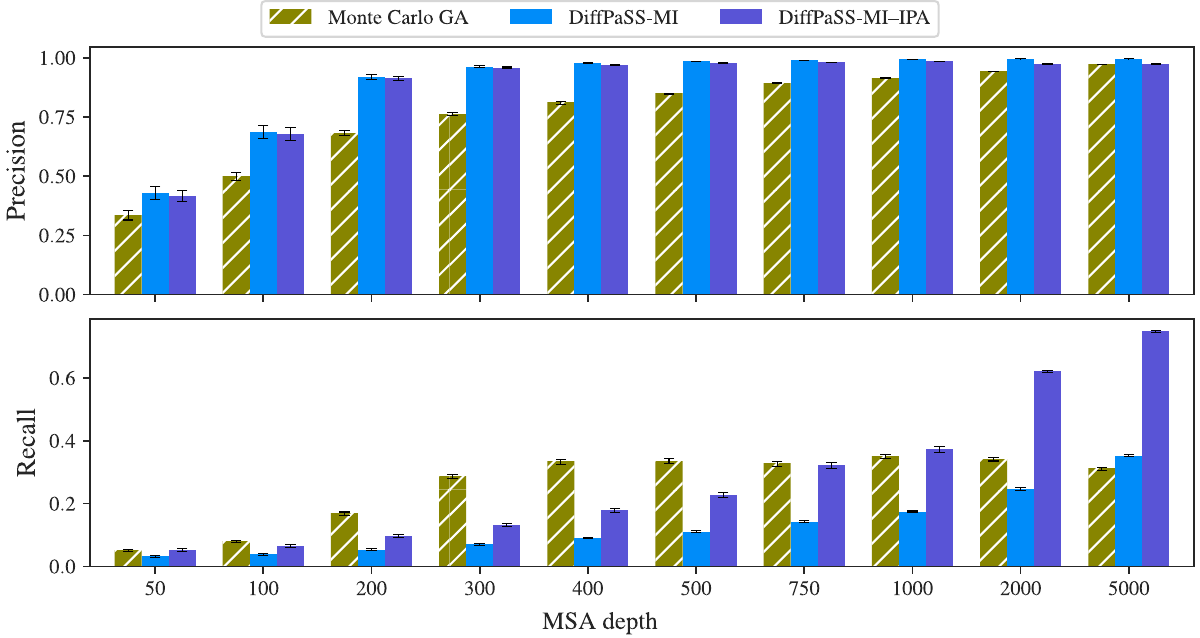}
    \caption{\textbf{Robust pairs.} Precision and recall for the robust pairs found by the Monte Carlo GA plus MI-IPA method (using 20 independent GA runs, as in \cite{Gandarilla23}), DiffPaSS-MI, and DiffPaSS-MI--IPA with $N_\mathrm{IPA} = 3$ for several MSA depths, on the HK-RR dataset. Same MSAs as in \cref{fig:hk-rr_results_comparison}.}
    \label{fig:precision_recall}
\end{figure}

We noticed that, even when the starting set $\mathcal{F}_{\mathrm{AB}}$ of fixed pairs is empty, some pairs are matched by all the $N_\mathrm{steps}$  hard permutations $\pi$ explored by DiffPaSS (see step \ref{step:hard_perm} in Supplementary material Section \ref{initialization_bootstrap}).
We call these \emph{robust pairs}, and notice that they tend to have high precision, see \cref{fig:precision_recall}.
This suggests that they can be used as the starting set $\mathcal{F}_{\mathrm{AB}}$ of fixed pairs in a further run of DiffPaSS.
We call this procedure DiffPaSS-IPA (``Iterative Pairing Algorithm'', following \citet{Bitbol16, Bitbol18}).
It can be iterated several times, enlarging the set of robust pairs after each iteration, or stopping if no further robust pairs are found.
The final output of DiffPaSS-IPA is the hard permutation with lowest observed loss across all IPA runs.
We use $N_\mathrm{IPA} = 3$ iterations throughout.

\subsection{Independence of number of Sinkhorn normalizations}
\label{sec:Theorem_Sinkhorn_normalizations}
Let $S^\ell$ (for an integer $\ell > 0$) be the $\ell$-truncated Sinkhorn operator defined in \cref{eq:sinkhorn_op}.
In this section we prove that, for any $\ell > 1$, all first-order derivatives of $S^\ell$, when evaluated at the zero matrix, are equal to the corresponding derivatives of $S^{\ell = 1}$.

Let $\mathcal{R}$ (resp.\ $\mathcal{C}$) be the row-wise (resp.\ column-wise) matrix normalization operator on $D \times D$ matrices.
Denote the partial derivative operator with respect to the $(i, j)$-th matrix entry by $\partial_{ij}$, and the $(k, l)$-th matrix component of a matrix-valued operator $\mathcal{O}$ by $\mathcal{O}_{kl}$.
Furthermore, let $\mathcal{T}^* = \mathcal{C} \circ \mathcal{R}$.
Let $\bm{1}_{\mathrm{mat}}$ denote the $D \times D$ matrix whose entries are all equal to $1$.
Given the definition of the Sinkhorn operator in \cref{eq:sinkhorn_op}, and since the componentwise exponential of the zero matrix is a matrix of ones, it suffices, for our purposes, to show that
\begin{equation}
    [\partial_{ij} (\mathcal{T}^* \circ \mathcal{T}^*)_{kl}](\bm{1}_{\mathrm{mat}}) = [\partial_{ij} \mathcal{T}^*_{kl}](\bm{1}_{\mathrm{mat}})
    \label{eq:tcirct_equals_t}
\end{equation}
for all $i, j, k, l = 1, \ldots, D$.
Indeed, we will prove that this actually holds when both sides of \cref{eq:tcirct_equals_t} are evaluated at $\mu \bm{1}_{\mathrm{mat}}$, for any real number $\mu > 0$.

\begin{proof}[Proof of \cref{eq:tcirct_equals_t}]
Let $\bm{X}$ denote a $D \times D$ matrix with positive entries.
We begin by noting that 
\begin{equation}
    [\partial_{ij} \mathcal{R}_{kl}](\bm{X}) = \delta_{ik} \, [\partial_{j} \mathcal{T}_l](\bm{X}_{i \cdot}) \quad \textrm{and} \quad [\partial_{ij} \mathcal{C}_{kl}](\bm{X}) = \delta_{jl} \, [\partial_{i} \mathcal{T}_k](\bm{X}_{\cdot j})\,,
    \label{eq:interm}
\end{equation}
where $\delta$ denotes the Kronecker delta, $\mathcal{T}$ the normalization operator for $D$-dimensional vectors, and $\bm{X}_{i \cdot}$ (resp.\ $\bm{X}_{\cdot j}$) the $i$-th row (resp.\ $j$-th column) of $\bm{X}$.

Let $\bm{x}$ denote a $D$-dimensional vector. The partial derivatives of the components of $\mathcal{T}$, evaluated at $\bm{x}$, are given by
\begin{equation}
    \partial_{\alpha} \mathcal{T}_{\beta} (\bm{x}) = \frac{\partial}{\partial \alpha} \frac{x_\beta}{\sum_{\gamma} x_\gamma} = \frac{\delta_{\alpha \beta}}{\sum_{\gamma} x_\gamma} - \frac{x_\beta}{(\sum_{\gamma} x_\gamma)^2}.
    \label{eq:deriv}
\end{equation}
Let $\bm{1}$ denote the $D$-dimensional vector whose entries are all equal to $1$.
Applying \cref{eq:deriv} to $\bm{x}=\mu \bm{1}$ yields, for any real number $\mu > 0$,
\begin{equation}
    \partial_{\alpha} \mathcal{T}_{\beta} (\mu \bm{1}) = \frac{1}{D \mu} (\delta_{\alpha\beta} - 1 / D) = \frac{1}{D \mu} \bm{\Delta}_{\alpha\beta},
    \label{eq:delta}
\end{equation}
where we have defined the square matrix $\bm{\Delta} = \mathrm{Id} - \bm{1}_\mathrm{mat} / D$, with $\mathrm{Id}$ denoting the $D \times D$ identity matrix.
We remark for later that $\bm{1}_\mathrm{mat} / D$ is idempotent (i.e., its matrix square is itself), and that therefore so is $\bm{\Delta}$.

Using the chain rule and the fact that $\mathcal{R}(\mu \bm{1}_\mathrm{mat}) = \bm{1}_\mathrm{mat} / D$ for any $\mu > 0$, we can write
\begin{align}
    [\partial_{ij} \mathcal{T}^*_{kl}] (\mu \bm{1}_\mathrm{mat}) &= \sum_{m, n} [\partial_{mn} \mathcal{C}_{kl}] (\bm{1}_\mathrm{mat} / D) \, [\partial_{ij} \mathcal{R}_{mn}] (\mu \bm{1}_\mathrm{mat}) \nonumber \\
    &= \sum_{m, n} \delta_{nl} \bm{\Delta}_{mk} \, \frac{\delta_{im}}{D \mu} \bm{\Delta}_{jn} \nonumber \\
    &= \frac{1}{D \mu} \bm{\Delta}_{ik} \bm{\Delta}_{jl}\,, \label{eq:one_deriv_norm}
\end{align}
where we used \cref{eq:interm} and \cref{eq:delta} to obtain the second line.

We now compute
\begin{align*}
    [\partial_{ij} (\mathcal{T}^* \circ \mathcal{T}^*)_{kl}] (\mu \bm{1}_\mathrm{mat}) &= \sum_{mn} [\partial_{mn} \mathcal{T}^*_{kl}](\bm{1}_\mathrm{mat} / D) \, [\partial_{ij} \mathcal{T}^*_{mn}](\mu \bm{1}_\mathrm{mat}) \\
    &= \frac{1}{D \mu} \sum_{mn} \bm{\Delta}_{mk} \bm{\Delta}_{nl} \bm{\Delta}_{im} \bm{\Delta}_{jn} \\
    &= \frac{1}{D \mu} (\bm{\Delta}^2)_{ik} (\bm{\Delta}^2)_{jl} \\
    &= [\partial_{ij} \mathcal{T}^*_{kl}] (\mu \bm{1}_\mathrm{mat}),
\end{align*}
where the last equality follows from the idempotency of $\bm{\Delta}$ and from \cref{eq:one_deriv_norm}.
When $\mu = 1$, this proves \cref{eq:tcirct_equals_t}.
\end{proof}

\section{Datasets}
\label{sec:Datasets}

\paragraph*{Benchmark prokaryotic datasets.}
We developed and tested DiffPaSS using joint MSAs extracted from a dataset composed of 23,632 cognate pairs of histidine kinases (HK) and response regulators (RR) from the P2CS database~\citep{Barakat09,Barakat11}, paired using genome proximity, and previously described in~\citet{Bitbol16,Bitbol18}. 
Our focus is on pairing interaction partners among paralogs within each species.
Pairing is trivial for a small number of species comprising only one pair of sequences. Hence, these species were discarded from the dataset.
The average number of pairs per species in the resulting dataset is $11.0$.

From this benchmark dataset of known interacting pairs, we extract paired MSAs of average depth 50, 100, 200, 300, 400, 500, 750 or 1000, constructed by selecting all the sequences of randomly sampled species from the full dataset.
Each depth bin contains at least $200$ MSAs.
More precisely, for a target MSA depth $\overline{D} = 50$, $100$, $200$, $300$, $400$, $500$, $750$ or $1000$, we add randomly sampled complete species one by one; if the first $m$ species (but no fewer) give an MSA depth $D \geq 0.9 \overline{D}$, and the first $n \geq m$ species (but no more) give $D \leq 1.1 \overline{D}$, then we select the first $k$ species in our final MSA, with $k$ picked uniformly at random between $m$ and $n$.

\paragraph*{Eukaryotic complexes.} \label{supp_datasets_eukaryotic}

We consider 15 heteromeric eukaryotic targets whose structures are not in the training set of AFM with v2 weights, already considered in \citet{Lupo2024}.

\paragraph*{T-cell receptor paired CDR3$\alpha$-CDR3$\beta$ data.}
\label{supp_datasets_tcr}
We downloaded the full VDJDb database \citep{goncharov_vdjdb_2022} and removed all entries where only a single TCR chain is available.
For each epitope, we removed duplicate TCRs (defined at the level of $\alpha$/$\beta$) and retained only epitopes for which at least 100 and no more than 10,000 sequences were available.
Patient and study metadata from the database is used to define ``groups'' that permutations are to be restricted to, and that play the part of species in our paralog matching problem.
Our final dataset contains $22$ sets of CDR3$\alpha$-CDR3$\beta$ sequence collections of highly variable total size (ranging from 103 to 1894 sequence pairs) and mean group size, for which the ground-truth matchings are known.

\section{General points on AlphaFold-Multimer (AFM)} \label{sec:generalities_AFM}

For all structure prediction tasks, we use the five pre-trained AFM models with v2 weights \citep{evans2021protein}.
We use full genomic databases and code from release v2.3.1 of the official implementation in \url{https://github.com/deepmind/alphafold}.
We use no structural templates, and perform 3 recycles for each structure, without early stopping.
We relax all models using AMBER.

We pair the same subset of pairable sequence retrieved by AFM as \citet{Lupo2024}, and refer to \citet[Table S1]{Lupo2024} for details on the pairable MSAs.
For all structures, we use the query protein pair as fixed context for DiffPaSS.

The AFM confidence score is defined as $0.8 \cdot \mathrm{iptm} + 0.2 \cdot \mathrm{ptm}$, where iptm is the predicted TM-score in the interface, and ptm the predicted TM-score of the entire complex \citep{evans2021protein}.

\newpage

\section{Supplementary figures}

\begin{figure}[h!]
    \centering
    \includegraphics[width=\textwidth]{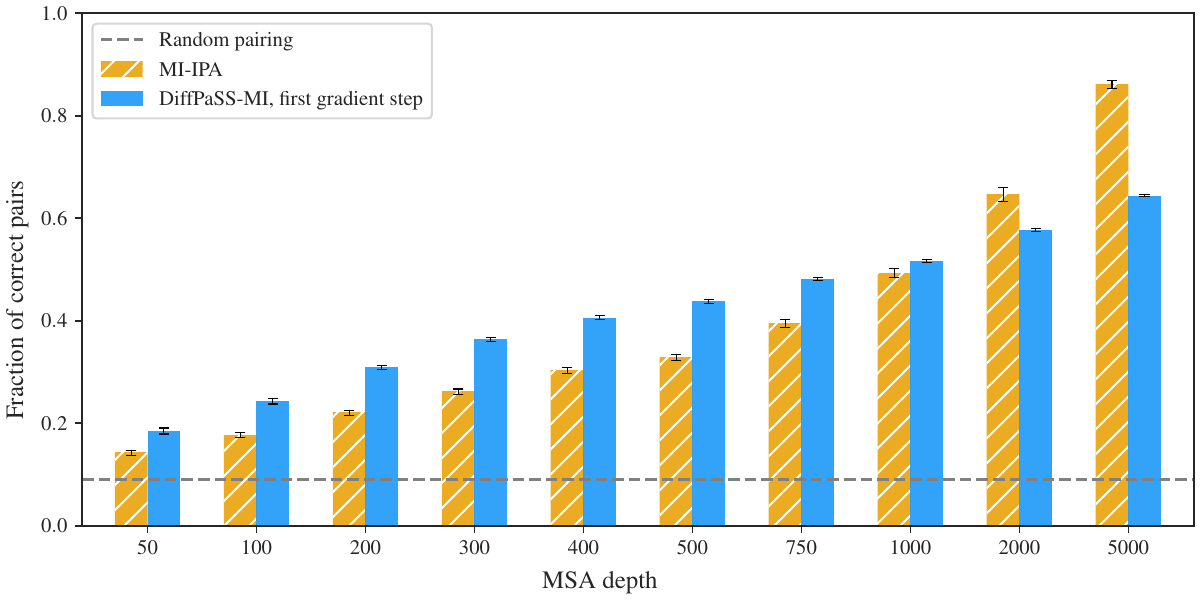}
    \caption{\textbf{DiffPaSS pairing performance after the first gradient step.} Pairing performance by DiffPaSS is shown after the first gradient step and compared to MI-IPA~\cite{Bitbol18} on HK-RR MSAs with various depths. Same as in \cref{fig:hk-rr_results_comparison} but restricting to the first gradient step. Here, our procedure, after only one gradient step, is competitive with MI-IPA \citep{Bitbol18}, for approximate maximization of the same discrete score $\mathscr{S}_\mathrm{MI}$.}
    \label{fig:hk-rr_precision_first_gradient_step}
\end{figure}

\begin{figure}[h!]
    \centering
    \includegraphics[width=\textwidth]{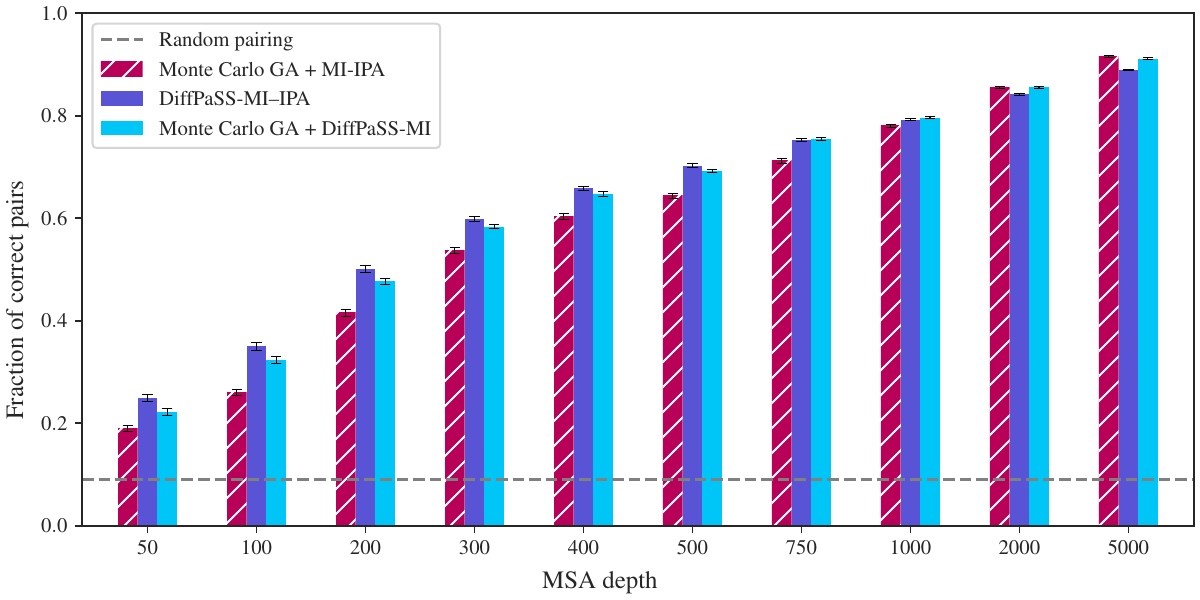}
    \caption{\textbf{Combining DiffPaSS-MI and Monte Carlo GA.} The performance of DiffPaSS-MI combined with Monte Carlo GA is evaluated on the same MSAs as in \cref{fig:hk-rr_results_comparison}, starting from the same robust pairs identified by Monte Carlo GA. 
    For comparison, we reproduce the bars showing the performance of DiffPaSS-MI--IPA alone, and of MI-IPA combined with Monte Carlo GA, from  \cref{fig:hk-rr_results_comparison}. }
    \label{fig:ga_plus_diffpass}
\end{figure}

\begin{figure}[h!]
    \centering
    \includegraphics[width=\textwidth]{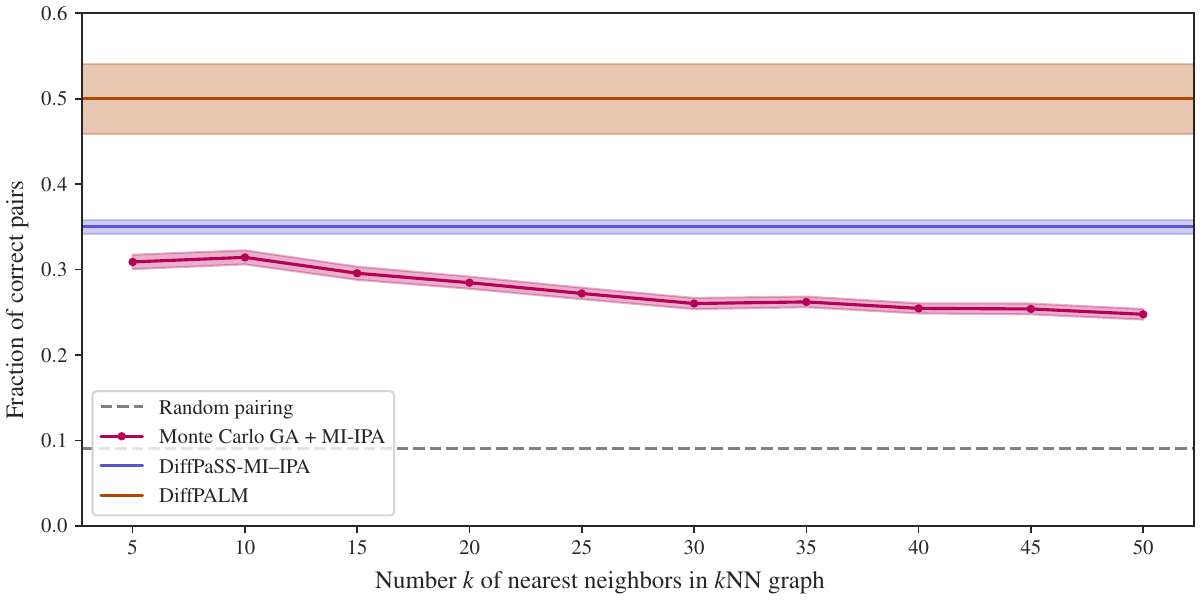}
    \caption{\textbf{Detailed comparison of MI-IPA combined with Monte Carlo GA with DiffPaSS-MI--IPA and DiffPALM, for shallow HK-RR MSAs.} For the same HK-RR MSAs of depth 100 as in \cref{fig:hk-rr_results_comparison}, we study the impact of the number $k$ of nearest neighbors in the $k$NN graph considered for Monte Carlo GA, see~\citep{Gandarilla23}, on the performance of MI-IPA combined with Monte Carlo GA~\cite{Gandarilla23}. The performances of DiffPaSS-MI--IPA and of DiffPALM for the same datasets are shown for comparison. Recall that they do not depend on $k$ since the $k$NN graph is not used in these methods.}
    \label{fig:gaipa_knn_test}
\end{figure}

\begin{figure}[h!]
    \centering
    \includegraphics[width=\textwidth]{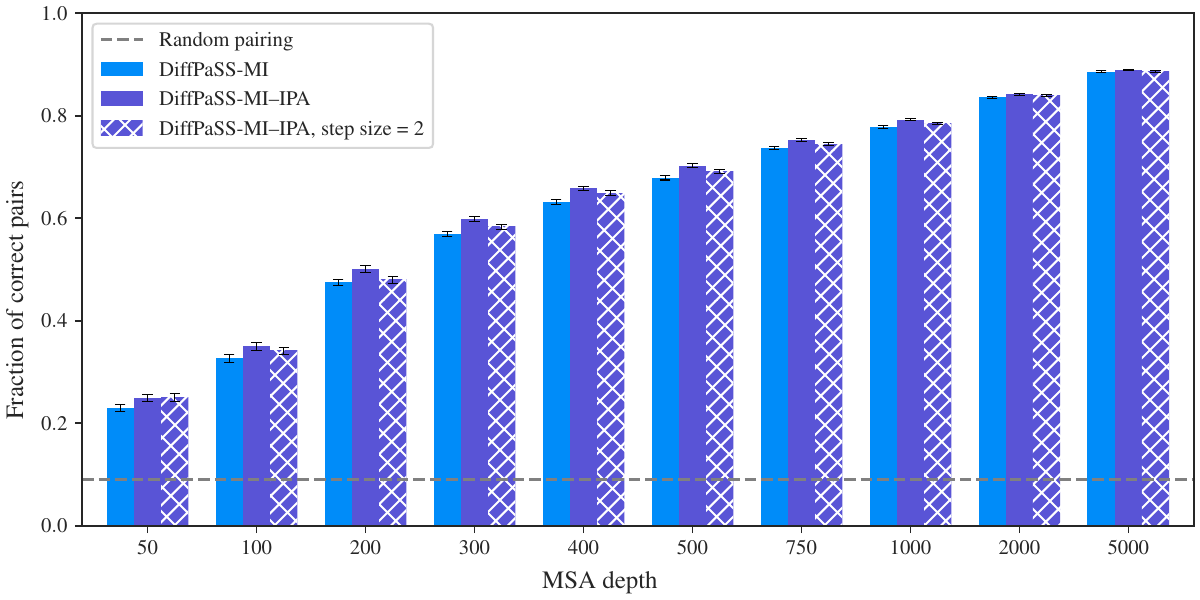}
    \caption{\textbf{Impact of the step size $\bm{\Delta n}$ on DiffPaSS-MI--IPA.} The performance of DiffPaSS-MI--IPA with step size 2 (i.e.\ when increasing by 2 the number of fixed pairs at each optimization run of DiffPaSS) is compared to that of DiffPaSS-MI--IPA with step size 1 (our default choice), and to that of DiffPaSS-MI, for HK-RR MSAs with different depths. Results for the latter two methods, shown for comparison, are identical to those in \cref{fig:hk-rr_results_comparison}. The MSAs and the performance metrics used are the same as in \cref{fig:hk-rr_results_comparison}.}
    \label{fig:step_size_2}
\end{figure}

\begin{figure}[h!]
    \centering
    \includegraphics[width=\textwidth]{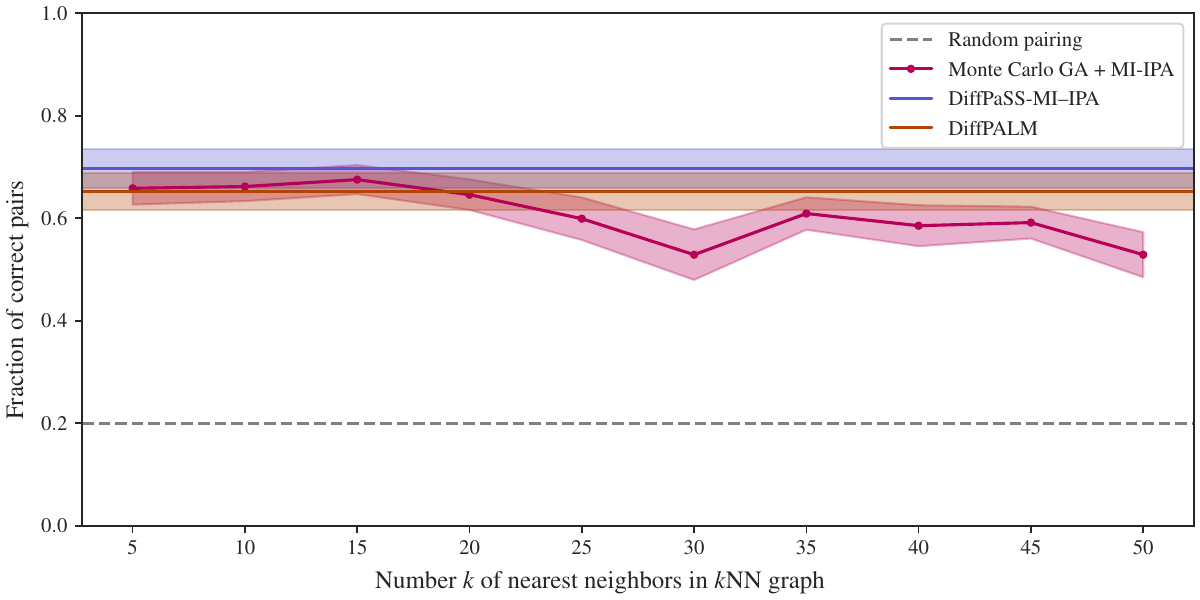}
    \caption{\textbf{Comparison of DiffPaSS-MI--IPA with DiffPALM and with MI-IPA combined with Monte Carlo GA for MALG-MALK.} Same as \cref{fig:gaipa_knn_test}, but for the same MALG-MALK MSAs of depth 100 as in~\citet{Lupo2024}.}
    \label{fig:malg-malk}
\end{figure}

\begin{figure}[h!]
	\centering
	\includegraphics[width=\textwidth]{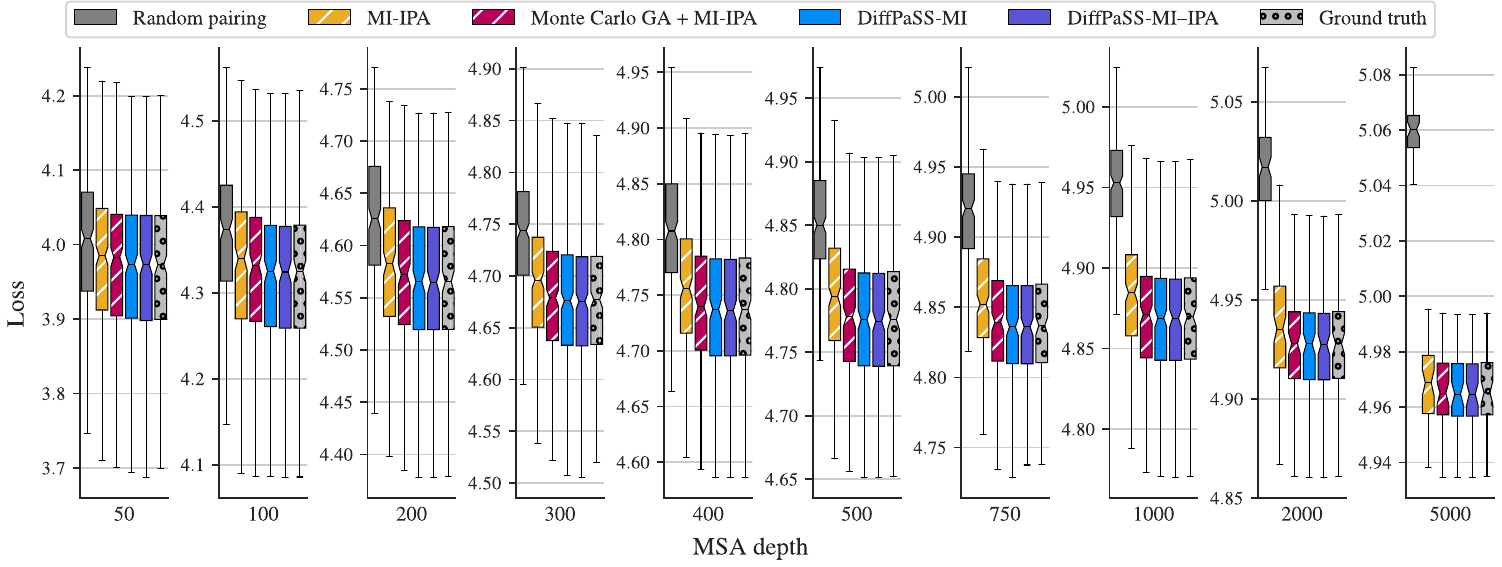}
	\caption{\textbf{Inter-chain two-body entropy losses for random, predicted and ground-truth pairings.} The distributions of the inter-chain two-body entropy losses are shown for several methods and several MSA depths, on the HK-RR dataset. Same MSAs as in \cref{fig:hk-rr_results_comparison}.}
	\label{fig:hk-rr_2be_comparison}
\end{figure}

\begin{figure}[h!]
    \centering
    \includegraphics[width=0.95\textwidth]{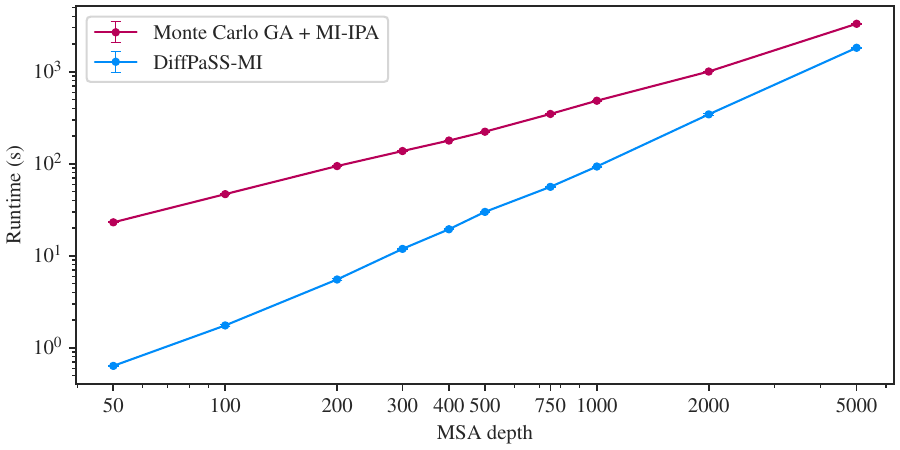}
    \caption{\textbf{Computational runtime.} Runtime comparison between DiffPaSS-MI and Monte Carlo GA + MI-IPA \citep{Gandarilla23}, on the same HK-RR MSAs as in \cref{fig:hk-rr_results_comparison}.
    DiffPaSS was implemented in \texttt{PyTorch} v2.2.1, and run on an NVIDIA\textregistered\ GeForce RTX\textsuperscript{TM} A6000 GPU.
    We used the original implementation of the combined Monte Carlo GA and MI-IPA method, in the Julia programming language (Julia v1.10.0).
    All 20 GA replicates were run in parallel on separate Intel\textregistered\ Xeon\textregistered\ Platinum 8360Y CPUs running at 2.4\,GHz and then MI-IPA was run on one of these CPUs.}
    \label{fig:hk-rr_diffpass_ga-ipa_runtime_comparison}
\end{figure}

\begin{figure}[htbp]
    \centering
    \includegraphics[width=0.95\textwidth]{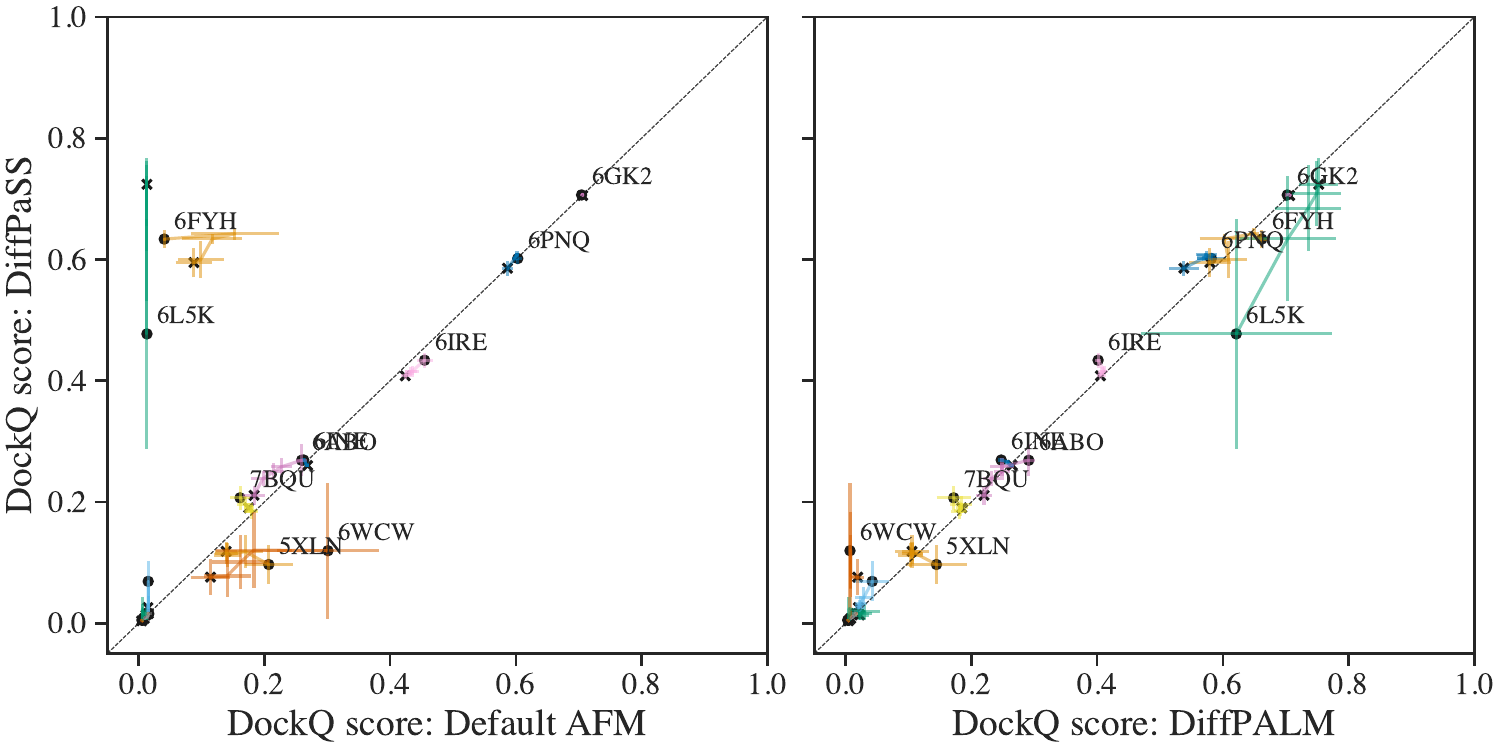}
    \caption{\textbf{Performance of structure prediction by AFM using different MSA pairing methods.} We report the performance of AFM, in terms of DockQ scores, for the 15 complexes considered in \cite{Lupo2024}, using different pairing methods on the same MSAs. Left panel: DiffPaSS versus default AFM pairing. Right panel: DiffPaSS versus DiffPALM. For each complex, AFM is run five times, and the five top predicted structures by AFM confidence are considered each time, yielding 25 predicted structures in total. For each complex, we show ``trajectories" of performance starting from the top-confidence predicted structure (black circular marker) and ending with all predicted structures up to and including the fifth one (black cross marker). Results are averaged over the 5 runs and standard errors are shown as error bars. Points with DockQ below $0.1$ are not labelled with their PDB ID for graphical reasons. 
    }
    \label{fig:dockq_comparison}
\end{figure}

\begin{figure}[h!]
    \centering
    \includegraphics[width=\textwidth]{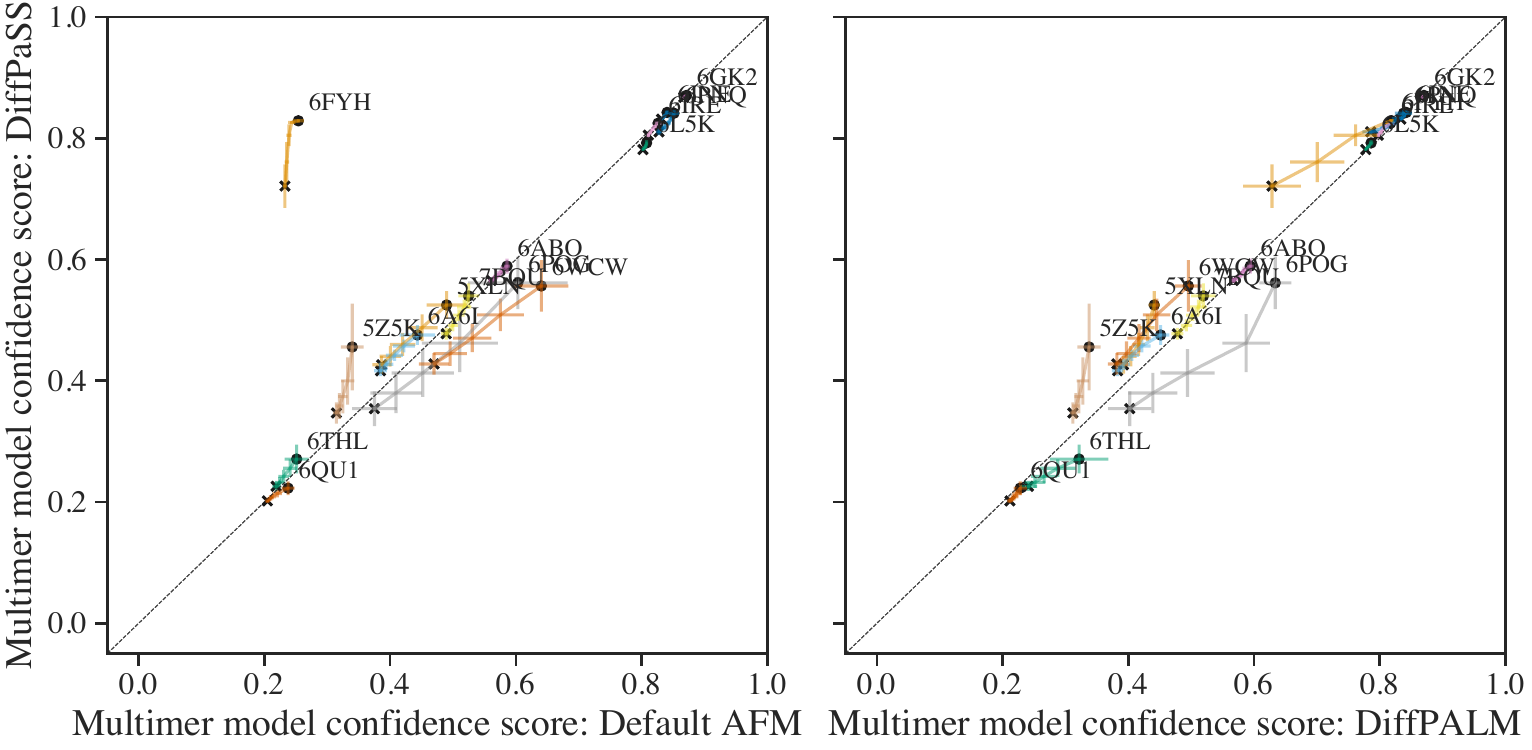}
    \caption{\textbf{AlphaFold-Multimer confidence scores on eukaryotic complexes using different MSA pairing methods.} Same comparisons as in \cref{fig:dockq_comparison}, but showing the confidence scores for structure prediction by AFM instead of DockQ scores. See Supplementary material Section \ref{sec:generalities_AFM} for a definition of this confidence score.
    Results for Default AFM and DiffPALM-based pairing are as in \citet[Fig.\ S5]{Lupo2024}.
}
    \label{fig:afm_confidences}
\end{figure}

\begin{figure}[h!]
    \centering
    \includegraphics[width=0.7\textwidth]{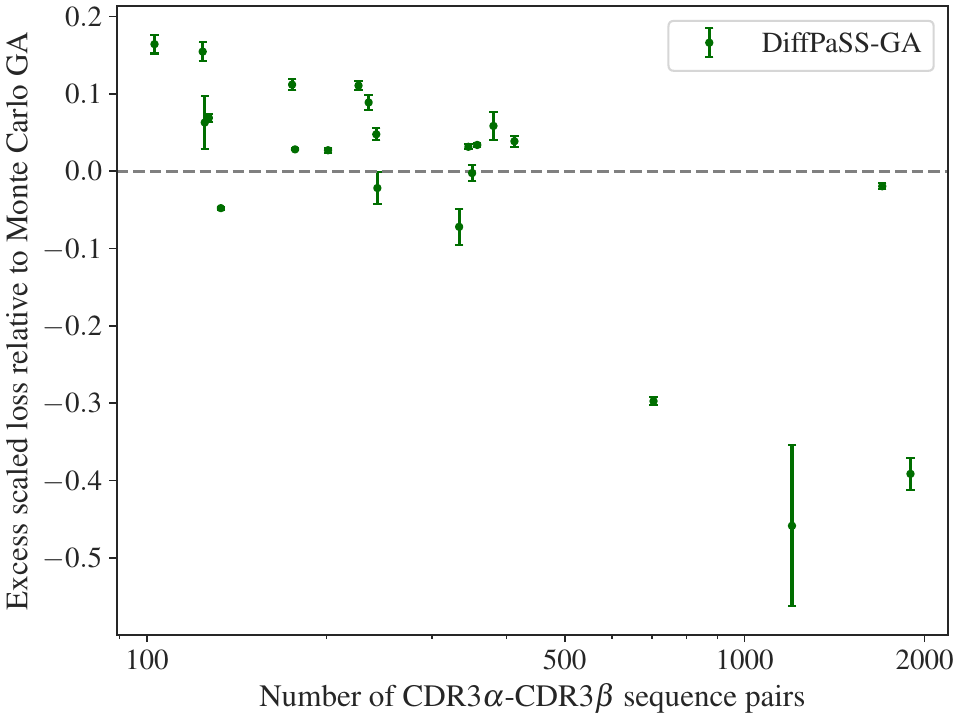}
    \caption{\textbf{DiffPaSS optimization and pairing quality on the CDR3$\alpha$-CDR3$\beta$ dataset.} (Normalized) graph alignment scores per epitope were computed in the paired CDR3$\alpha$-CDR3$\beta$ dataset described in Supplementary material Section \ref{sec:Datasets} with both DiffPaSS-GA and the Monte Carlo GA from \citet{Gandarilla23}. We plot the difference between the loss obtained by DiffPaSS-GA and by Monte Carlo GA versus the number of pairs for each epitope. Negative values indicate that Monte DiffPaSS-GA achieves a more successful optimization.
    Each point represents one of the epitopes in the dataset. Markers show mean values across 20 runs, and error bars are defined as $\sqrt{\sigma_\textrm{MC}^2 + \sigma_\textrm{DiffPaSS}^2}$, where $\sigma_\textrm{MC}$ (resp.\ $\sigma_\textrm{DiffPaSS}$) is the standard deviation obtained by Monte Carlo GA (resp.\ DiffPass-GA).
    Graph alignment scores are as defined in \citep{Gandarilla23}, but normalized by the number of CDR3$\alpha$ and CDR3$\beta$ sequences to pair in each case.}\label{fig:CDR_diffpass_vs_simulated_annealing}
\end{figure}

\end{document}